\documentclass[twocolumn,superscriptaddress,aps,preprintnumbers,amsmath,amssymb,sort&compress,nofootinbib
]{revtex4-1}
 
\usepackage{amsfonts,amsmath,amssymb,ascmac,bm,tensor}
\usepackage{fnpct} 
\usepackage{comment}
\usepackage{ifpdf}
\usepackage{slashed}
\usepackage{color}
\usepackage[mathscr]{eucal}
\usepackage{fourier}
\usepackage[utf8]{inputenc}

\usepackage{cancel}

\ifpdf
  \usepackage{graphicx}     
  \usepackage[bookmarksopen,colorlinks=true,linkcolor=bblue,citecolor=bblue,urlcolor=ppink]{hyperref}
\else     
  \usepackage[dvipdfmx]{graphicx}     
  \usepackage[dvipdfmx,bookmarksopen,colorlinks=true,linkcolor=bblue,citecolor=ppink,urlcolor=darkred]{hyperref}
\fi

\definecolor{red}{rgb}{1,0,0}
\definecolor{darkred}{rgb}{0.6,0,0}
\definecolor{darkgreen}{rgb}{0.992447,0.623778,0.034597}
\definecolor{ppink}{rgb}{1,0.4,0.4}
\definecolor{bblue}{rgb}{0.284602,0.317763,0.963947}
\definecolor{purple}{rgb}{0.5 ,0, 0.7}

\newcommand{\vev}[1]{ \left< {#1} \right> }

\newcommand{\dd}{\mathrm{d}}

\newcommand{\NL}{\text{NL} }
\newcommand{\R}{\text{R} }

\newcommand{\ee}{\mathrm{e}}

\newcommand{\eqf}{{\text{eq},1}}
\newcommand{\eqs}{{\text{eq},2}}
\newcommand{\PBH}{{\text{PBH}}}

\newcommand{\calP}{\mathcal{P}}

\newcommand{\sma}{\ell}  
\newcommand{\cmd}{x}  
\newcommand{\efinal}{e_\text{final}}  
\newcommand{\enn}{e_\text{NN}}  

\makeatletter
\newcommand\footnoteref[1]{\protected@xdef\@thefnmark{\ref{#1}}\@footnotemark}
\makeatother

\allowdisplaybreaks[1]

\begin{document}


\title{
Gravitational Wave Production right after a Primordial Black Hole Evaporation
}
\affiliation{ICRR, University of Tokyo, Kashiwa, 277-8582, Japan}
\affiliation{DESY, Notkestra{\ss}e 85, D-22607 Hamburg, Germany}
\affiliation{Center for Theoretical Physics of the Universe, Institute for Basic Science (IBS),
  Daejeon, 34126, Korea}
\affiliation{T. D. Lee Institute and School of Physics and Astronomy, Shanghai Jiao Tong University, 800 Dongchuan Rd, Shanghai 200240, China}
\affiliation{Kavli IPMU (WPI), UTIAS, University of Tokyo, Kashiwa, 277-8583, Japan}
\author{Keisuke Inomata}
\affiliation{ICRR, University of Tokyo, Kashiwa, 277-8582, Japan}
\affiliation{Kavli IPMU (WPI), UTIAS, University of Tokyo, Kashiwa, 277-8583, Japan}
\author{Masahiro Kawasaki}
\affiliation{ICRR, University of Tokyo, Kashiwa, 277-8582, Japan}
\affiliation{Kavli IPMU (WPI), UTIAS, University of Tokyo, Kashiwa, 277-8583, Japan}
\author{Kyohei Mukaida}
\affiliation{DESY, Notkestra{\ss}e 85, D-22607 Hamburg, Germany}
\author{Takahiro Terada}
\affiliation{Center for Theoretical Physics of the Universe, Institute for Basic Science (IBS),
  Daejeon, 34126, Korea}
\author{Tsutomu T.~Yanagida}
\affiliation{T. D. Lee Institute and School of Physics and Astronomy, Shanghai Jiao Tong University, 800 Dongchuan Rd, Shanghai 200240, China}
\affiliation{Kavli IPMU (WPI), UTIAS, University of Tokyo, Kashiwa, 277-8583, Japan}

\begin{abstract}
\noindent
We discuss the footprint of evaporation of primordial black holes (PBHs) on stochastic gravitational waves (GWs) induced by scalar perturbations. We consider the case where PBHs once dominated the Universe but eventually evaporated before the big bang nucleosynthesis. 
The reheating through the PBH evaporation could end with a sudden change in the equation of state of the Universe compared to the conventional reheating caused by particle decay.
We show that this ``sudden reheating'' by the PBH evaporation enhances the induced GWs, whose amount depends on the length of the PBH-dominated era and the width of the PBH mass function. 
We explore the possibility to constrain the primordial abundance of the evaporating PBHs by observing the induced GWs.  
We find that the abundance parameter $\beta \gtrsim 10^{-5} \text{ -- }10^{-8}$ for $\mathcal{O}(10^3 \text{ -- } 10^5) \, \text{g}$ PBHs can be constrained by future GW observations if the width of the mass function is smaller than about a hundredth of the mass. 
\end{abstract}

\date{\today}
\maketitle
\preprint{IPMU 20-0029}
\preprint{DESY 20-042}
\preprint{CTPU-PTC-20-05}

\section{Introduction}
\label{sec:intro}

Primordial black holes (PBHs)~\cite{Hawking:1971ei,Carr:1974nx,Carr:1975qj} have been gathering interests for many years despite its lack of observational evidence.
This is particularly because of their rich phenomenology in Cosmology for a wide range of their masses.
Depending on their mass range, PBHs that survive until today ($\gtrsim 10^{15}$\,g) could explain dark matter~\cite{Carr:2016drx,Inomata:2017okj,Inomata:2017vxo}, 
the gravitational waves (GWs) from the black hole (BH) mergers detected by LIGO/Virgo~\cite{Bird:2016dcv,Clesse:2016vqa,Sasaki:2016jop}, the microlensing events in the OGLE data~\cite{mroz2017no,Niikura:2019kqi} (see also Ref.~\cite{Sasaki:2018dmp} for a review), and cosmic structures such as the seeds of supermassive BHs~\cite{Kawasaki:2012kn,Carr:2018rid}.
On the other hand, tiny PBHs that had evaporated before the big bang nucleosynthesis (BBN) ($\lesssim 10^{9}$\,g) could generate the baryon asymmetry~\cite{Toussaint:1978br,Turner:1979zj,Turner:1979bt,Barrow:1990he,Majumdar:1995yr,Upadhyay:1999vk,Dolgov:2000ht,Bugaev:2001xr,Baumann:2007yr,Fujita:2014hha, Hamada:2016jnq}, produce dark matter particles~\cite{Fujita:2014hha,Lennon:2017tqq,Morrison:2018xla}, and relax the Hubble tension~\cite{Hooper:2019gtx,Nesseris:2019fwr,Lunardini:2019zob}.

Following the first detection of GWs from $\mathcal{O} (10) M_\odot$ black-hole mergers, observational constraints on PBHs for a wide range of their masses 
have been reconsidered.
For PBHs with $M_\text{PBH} > 10^9$\,g, there are a number of constraints on the abundance of PBHs, which are considered as robust~\cite{Carr:2009jm, Sasaki:2018dmp, Carr:2020gox}.
Even evaporating PBHs in this mass range can be constrained by the null detection of the extragalactic or galactic gamma rays from Hawking radiation and by their effects on the BBN and the cosmic microwave background (CMB)~\cite{Carr:2009jm, Laha:2019ssq, Dasgupta:2019cae,Carr:2020gox}. 

However, conservative constraints on the tiny PBHs with $M_\text{PBH} < 10^9$\,g are still lacking.\footnote{\label{fn:pbh_catalyst}
	In Refs.~\cite{Gregory:2013hja,Burda:2015yfa,Burda:2015isa,Burda:2016mou,Tetradis:2016vqb}, the possibility of evaporating PBHs as a catalyst of vacuum decay was pointed out, which could be used to constrain the PBHs abundance assuming our electroweak vacuum is metastable~\cite{Kohri:2017ybt}.
	However, this bound depends on the UV completion or a precise value of the  top Yukawa, and moreover, there are subtleties because the analysis made so far also predicts the bubble nucleation whose typical size is much larger than the PBH radius, implying a suppression factor from this effect is missing~\cite{Gorbunov:2017fhq,Mukaida:2017bgd}.
}
The constraints are so weak that it is even possible that the PBHs dominate the energy density of the Universe before they evaporate.  
There exist few attempts to probe this mass regime in the literature.
One could examine the PBH abundance through its stable relics~\cite{Carr:1994ar}, while whether or not a PBH leaves the relic requires dedicated studies of the quantum gravities and is still in debate~\cite{Chen:2014jwq}. 
Similarly, one could constrain the PBH abundance by the overproduction of dark matter through the PBH evaporation~\cite{Green:1999yh,Lemoine:2000sq,Khlopov:2004tn}.
However, it strongly depends on the nature of dark matter and cannot be applied directly to, e.g., axion or PBH dark matter. 
Also, if the PBH evaporation generates baryon asymmetry of the Universe~\cite{zeldovich1976charge,zeldovich1976possibility,Toussaint:1978br,Turner:1979zj,Turner:1979bt,Barrow:1990he,Majumdar:1995yr,Upadhyay:1999vk,Dolgov:2000ht,Bugaev:2001xr,Baumann:2007yr,Fujita:2014hha}, or if the baryon asymmetry is generated after the evaporation of PBHs, 
the constraint from the entropy production~\cite{zeldovich1976possibility} is not applicable.

Despite the difficulty of investigating the tiny PBHs, the production of them is predicted in the context of the hybrid inflation~\cite{GarciaBellido:1996qt} (see also Refs.~\cite{Clesse:2015wea,Kawasaki:2015ppx}), the inflation model with the Chern-Simons coupling between the inflaton and gauge fields~\cite{Linde:2012bt,Domcke:2020zez}, and the preheating after inflation~\cite{Martin:2019nuw,Martin:2020fgl}. 
In particular, the hybrid inflation is an attractive model in view of the see-saw mechanism and leptogenesis~\cite{Asaka:1999yd, Asaka:1999jb}. 
In this sense, the tiny PBHs could give us hints on not only the early Universe but also particle physics models.

In this paper, we shed light on the tiny PBHs in terms of GWs.
Once GWs are produced by some mechanism, they are not erased by the frictions with other matter species unlike the radiation perturbations, so GWs can be a good probe of the tiny PBHs. 
However, GWs emitted through the Hawking radiation~\cite{Anantua:2008am, Dolgov:2011cq, Dong:2015yjs} and mergers of the tiny PBHs~\cite{Zagorac:2019ekv} have very high frequencies.
This is because, for tiny PBHs, the Hawking temperature is high and the typical length scale of the PBH binary is short.
It is unlikely to detect such high-frequency GWs by the near-future GW observations (see also Appendix~\ref{app:various_GWs}).

Throughout this paper, we instead focus on the GWs induced by the scalar (curvature/density) perturbations which are related to the tiny PBHs and can be detected by the future observations. 
The scalar perturbations can be a source of GWs through their interactions appearing at the second order in perturbations~\cite{tomita1967non,Matarrese:1993zf,Matarrese:1997ay,Ananda:2006af,Baumann:2007zm,Saito:2008jc,Saito:2009jt}. 
The induced GWs have recently attracted a lot of attention~\cite{Inomata:2016rbd,Ando:2017veq,Espinosa:2018eve,Kohri:2018awv,Cai:2018dig,Bartolo:2018evs,Bartolo:2018rku,Unal:2018yaa,Byrnes:2018txb,Inomata:2018epa,Clesse:2018ogk,Cai:2019amo,Cai:2019jah,Wang:2019kaf,Ben-Dayan:2019gll,Tada:2019amh,Inomata:2019zqy,Inomata:2019ivs,Yuan:2019udt,Xu:2019bdp,Cai:2019elf,Lu:2019sti, Yuan:2019wwo, Chen:2019xse, Hajkarim:2019nbx,Ozsoy:2019lyy,Domenech:2019quo,Fu:2019vqc,Ota:2020vfn,Lin:2020goi,Ballesteros:2020qam} because the induced GWs can be used to investigate the small-scale ($k>1$\,Mpc$^{-1}$) perturbations, which are difficult to be accessed by CMB observations but can produce PBHs.

The induced GWs can be strong in the following two cases: (1) the primordial scalar perturbation is large, and (2) the scalar perturbation grows dynamically. 
In our scenario, the first case corresponds to the large scalar perturbation required to have a sizable amount of tiny PBHs.
The induced GWs are generated when the enhanced scalar perturbation responsible for the PBH formation enters the horizon.
Hence the peak frequency of the induced GWs has one to one correspondence to the PBH mass.
Because of this fact, the peak frequency for the tiny PBHs ($M_\text{PBH} < 10^{9}$\,g) is still high (see Appendix~\ref{app:various_GWs}).

The second case involves an early matter-dominated (eMD) era which precedes the standard radiation-dominated (RD) era.  During the eMD era, density perturbations grow on the subhorizon scales, and the gravitational potential (hence the source term of the induced GWs) does not decay on the subhorizon scales~\cite{Assadullahi:2009nf, Baumann:2007zm}.  
After the eMD era ends, the gravitational potential starts to oscillate due to the radiation pressure during the RD era.
During the reheating transition from the eMD era to the RD era (before their oscillation), the gravitational potentials on the subhorizon scales decay.
Hence, the amount of their decay depends on the time scale of the transition.
If the time scale of the reheating transition is sufficiently short, the gravitational potential on the subhorizon scales does not much decay before their fast oscillations since there is no time for it to decay. 
In this case, strong GWs are induced after the sudden reheating transition even if the power spectrum of the primordial curvature perturbations is almost scale invariant (i.e.~with no ad hoc enhancement in the initial condition)~\cite{Inomata:2019ivs}.  
Note that the fast oscillations of the gravitational potential in the RD era are caused by {\em sound waves} in the thermal bath, which is produced after the sudden disappearance (or ``demise'') of the matter field. 
For this reason, we call this mechanism the \textit{Poltergeist mechanism} for GW production.

Indeed, the sudden reheating transition can be realized in the PBH-dominating scenario.
A key property of evaporating PBHs in this context is that the evaporation process becomes faster and faster once it sets in because of the negative specific heat of a BH.  
If the PBHs come to dominate the Universe by the time when this explosive event happens, the equation of state for the Universe can change suddenly at the end of evaporation depending on how sharp the PBH mass function is. 
Since the PBH-dominated era behaves as an eMD era, the evaporation of the PBHs leads to a sudden transition from the eMD era to the RD era.
For this reason, we expect that tiny PBHs can trigger the Poltergeist mechanism: after the sudden evaporation of PBHs, the ``ghost'' of PBHs makes merry in the thermal bath producing strong GWs.
In addition, the induced GWs are enhanced at least for modes that enter the horizon by the end of the reheating.
This is a macroscopic wavelength of a fluid composed of many tiny PBHs, and therefore the induced GWs can have the frequencies lower than those of the other types of GWs mentioned above.  Thus, they can be detected by near-future GW observations if the enhancement is sufficiently large. 
For this reason, we focus on the GWs induced by the Poltergeist mechanism to investigate the tiny PBHs throughout this paper.

We study how large the enhancement of the induced GWs by the Poltergeist mechanism can be by taking into account the finite duration of the evaporation/reheating process. 
We also discuss novel observational consequences of the enhanced production of GWs from PBH evaporation. 
Assuming the (almost) scale-invariant power spectrum for the scalar perturbations $\calP_\zeta \sim 10^{-9}$ conservatively,
we estimate the spectrum of the enhanced GWs as a function of the PBH mass, its abundance (formation probability), and the width of the mass function both numerically and analytically. 
Utilizing such relations, we discuss prospective constraints on the primordial abundance of evaporating PBHs, which are accessible by future GW observations, such as LISA~\cite{Sathyaprakash:2009xs,Moore:2014lga,Audley:2017drz}, DECIGO~\cite{Seto:2001qf,Yagi:2011wg} and BBO~\cite{phinney2003big,Yagi:2011wg}.

This paper is organized as follows.
In Sec.~\ref{sec:pbh_era}, we review the PBH-dominated era. 
In particular, we see why the evaporation of PBHs ends with a sudden transition compared to the conventional reheating.
In Sec.~\ref{sec:evo_pertb}, we discuss the evolutions of scalar perturbations, focusing on the evolution of the gravitational potential, which is the source of the induced GWs, during the transition from the PBH-dominated era to the RD era.
In Sec.~\ref{sec:induced_gws}, we discuss the GWs induced by the scalar perturbations that have experienced the PBH-dominated era on subhorizon scales. 
Note that, until Sec.~\ref{sec:induced_gws}, to show the essential points of the Poltergeist mechanism, we assume the monochromatic PBH mass function for simplicity.
Then, we take into account the finite width of PBH mass function and discuss how it affects the induced GWs in Sec.~\ref{sec:width}.
In Sec.~\ref{sec:measure_pbh}, we discuss the prospective constraints on the PBH abundance from the future GW observations.
Finally, we conclude this paper in Sec.~\ref{sec:conclusions}.

Throughout this paper, we assume that the evaporation of PBHs does not leave any relics.
As a convention, we use the reduced Planck mass ($M_\text{Pl} \equiv 1/\sqrt{8\pi G}$) instead of the gravitational constant.
In addition, we use the word ``reheating'' to represent the reheating caused by the PBH evaporation (not the reheating caused by inflaton decay) unless otherwise noted.

\section{PBH-dominated era}
\label{sec:pbh_era}

The cosmological scenario we consider is as follows. After inflation and (p)reheating of the Universe caused by inflaton decay/annihilation, the Universe is filled with radiation. In this early radiation-dominated (eRD) era, a PBH is supposed to form when a rare and large perturbation mode enters the Hubble horizon. PBHs behave as non-relativistic matter, so they would eventually dominate the energy density of the Universe if they were stable.  Since PBHs are quantum-mechanically unstable due to the Hawking radiation, it depends on the initial abundance whether they dominate the Universe or not.  We are interested in the case where PBHs do dominate, and the condition for the domination is shortly reviewed below.  The PBH-dominated era serves as an eMD era.  It ends via the evaporation of PBHs.  Thus, the evaporation of the PBHs can be regarded as the reheating transition from the eMD era to the standard RD era.  The subsequent evolution of the Universe is the same as in the standard cosmological scenario.
In this scenario of PBH domination and evaporation, there are various mechanisms to produce GWs, which are reviewed in Appendix~\ref{app:various_GWs}.

We can easily generalize our discussion, e.g., to the cases of PBH formation in another matter-dominated (MD) era (e.g. during inflaton coherent oscillation) or PBH formation by phase transitions etc., but we do not do so here for simplicity.
In the rest of this section, we summarize various relations in the PBH-dominated era, which are useful in the subsequent sections.

\subsection{PBH evaporation}

Let us start with the governing equation of the PBH evaporation.
The mass of a PBH obeys~\cite{Hooper:2019gtx}
\begin{align}
 \frac{\dd M_\text{PBH}}{\dd t} &= - \frac{A}{M_\text{PBH}^2} \nonumber \\
 &= -7.6 \times 10^{16} \,\text{g}\, \text{s}^{-1}\, g_{\text{H}*}(T_\text{PBH}) \left( \frac{M_\text{PBH}}{10^4 \, \text{g}} \right)^{-2},
 \label{eq:m_pbh_evo}
\end{align}
where $A$ is given as 
\begin{align}
	A = \frac{\pi \, \mathcal G \, g_{H*}(T_\text{PBH}) M_\text{Pl}^4}{480}.
\end{align}
$\mathcal G \simeq 3.8$ is the gray-body factor and $T_\text{PBH}$ is the Hawking temperature of the PBH~\cite{Hawking:1974sw} 
\begin{align}
  T_\text{PBH} = \frac{M_\text{Pl}^2}{M_\text{PBH}} \simeq 1.05 \times 10^{9} \, \text{GeV} \, \left( \frac{M_\text{PBH}}{10^4 \, \text{g}} \right)^{-1}.
  \label{eq:t_pbh_m_pbh}
\end{align}
$g_{\text{H} *}(T_\text{PBH})$ is the spin-weighted degrees of freedom of the particles produced from the Hawking radiation with $T_\text{PBH}$, whose concrete value is given as~\cite{Hooper:2019gtx}
\begin{align}
  g_{\text{H}*} (T_\PBH) \simeq \begin{cases}
  108 & (T_\PBH \gg 100 \, \text{GeV} \leftrightarrow M_\PBH \ll 10^{11}\, \text{g} ) \\
  7 & ( T_\PBH \ll 1\, \text{MeV} \phantom{w} \leftrightarrow M_\PBH \gg 10^{16}\, \text{g} )
  \end{cases}.
\end{align}
The temperature dependence comes from the fact that the Hawking radiation cannot efficiently produce the particles heavier than the Hawking temperature ($m \gtrsim T_\text{PBH}$).

Solving Eq.~(\ref{eq:m_pbh_evo}), we can derive the time dependence of the PBH mass as 
\begin{align}
  M_\text{PBH} = \left(3A\right)^{1/3} \left(t_\text{eva} - t\right)^{1/3} \quad  \text{for}~~t \leq t_\text{eva}, \label{eq:mass-time_relation}
\end{align}
where the subscript ``eva'' indicates the value when the PBH completes the evaporation.
We will express $t_\text{eva}$ as temperature in Eq.~\eqref{eq:tr_pbh_mass}.
Since we focus on tiny PBHs with $M_\text{PBH} < 10^9\, \text{g}$ throughout this paper, we take $g_{\text{H}*} = 108$ and consider $A$ to be time-independent. 
Assuming the monochromatic PBH mass function, we can express the decay rate as
\begin{align}
  \Gamma \equiv - \frac{1}{M_\text{PBH}} \frac{\dd M_\text{PBH}}{\dd t} = \frac{1}{3\left(t_\text{eva} -t \right)},
  \label{eq:decay_rate}
\end{align}
where this decay rate is defined so that $\Gamma \rho_\PBH$ represents the energy flow from the PBHs to radiation per unit time and volume.
Note again that we discuss the effects of the finite width of PBH mass function in Sec.~\ref{sec:width}.
Now it is clear that,
in contrast to the perturbative decay of heavy particles, the decay rate grows towards the completion of the evaporation $t \to t_\text{eva}$, implying that this process is more sudden than the conventional reheating.

By using the relation $H=2/(3 t)$, one may express the evaporation time $t_\text{eva}$ as the reheating temperature, which is given by~\cite{Hooper:2019gtx}\footnote{In Ref.~\cite{Hooper:2019gtx}, the authors take $H \equiv \dd a/(a\,\dd t)=1/(2 t)$ at the end of the evaporation. However, we take $H=2/(3 t)$ at that time because we assume a PBH-dominated era before the evaporation. This is why the factor in Eq.~(\ref{eq:tr_pbh_mass}) is different from that in Ref.~\cite{Hooper:2019gtx}.}\footnote{
  We assume that the thermalization instantaneously occurs soon after the PBH evaporation.
  Since the typical momentum of the particles produced by the evaporation ($\sim T_\text{PBH}$) is much larger than the reheating temperature ($T_\R$), the thermalization occurs dominantly through the scatterings with small angles, which make the thermalization instantaneous in most cases~\cite{Harigaya:2013vwa}.
} 
\begin{align}
  \label{eq:tr_pbh_mass}
  T_\text{R} 
   \simeq 2.8 \times& 10^4 \, \text{GeV} \left( \frac{M_{\text{PBH,i}}}{10^4 \, \text{g}} \right)^{-3/2}  \nonumber \\
& \quad  \times\left( \frac{g_{\text{H}*}(T_\text{PBH})}{108} \right)^{1/2} \left( \frac{g_{*, \text{eva}}}{106.75} \right)^{-1/4},
\end{align}
where the subscript ``i'' represents the initial value (at the PBH production).
$g_\ast$ is the relativistic effective degrees of freedom, which should not be confused with $g_{\text{H}*}$.
The temperature dependence of $g_*$ is given in Refs.~\cite{Aghanim:2018eyx, Saikawa:2018rcs}.
For later convenience, we express the inverse horizon scale at the reheating as a function of reheating temperature~\cite{Inomata:2019ivs}:
\begin{align}
   k_\text{eva} = &4.7 \times 10^{11}\, \text{Mpc}^{-1} \left( \frac{g_{*,\text{eva}}}{106.75} \right)^{1/2}  \nonumber \\
   & \quad  \times \left(\frac{g_{s*,\text{eva}}}{106.75} \right)^{-1/3} \left( \frac{T_\text{R}}{2.8 \times 10^{4}\, \text{GeV} } \right),
  \label{eq:kr_tr} 
\end{align}
where $g_{s*}$ is the effective degrees of freedom for an entropy density.
Note that $g_{*,\text{eva}}$ and $g_{s*,\text{eva}}$ mean the values at $T = T_\text{R}$.
We have used the relation $k_\bullet = a_\bullet H_\bullet$ with $\bullet = \text{eva}, \text{eq}$ and taken the ratio $k_\text{eva} / k_\text{eq} = a_\text{eva} H_\text{eva} / (a_\text{eq} H_\text{eq})$ to derive this equation. 
The subscript ``eq'' means the value at the late-time matter-radiation equality ($z_\text{eq} \simeq 3400$).

\subsection{Condition for PBH domination}

The initial PBH mass is related to the temperature of the Universe at the PBH production as~\cite{Inomata:2017vxo}\footnote{
From the constraint on the tensor-to-scalar ratio ($r<0.065$~\cite{Aghanim:2018eyx}), the upper bound of the temperature after an inflaton decay is given as $T_\text{R,inf} < 6.7\times 10^{15} (g_*/106.75)^{-1/4}$\,GeV, where the instantaneous reheating occurring after the inflation era is assumed. 
  This means $T_i < 6.7\times 10^{15} (g_{*,i}/106.75)^{-1/4}$\,GeV, implying a lower bound on PBH mass of $M_{\text{PBH,i}} > 0.4$\,g $(\gamma/0.2)$.
}
\begin{align}
 M_{\text{PBH,i}} & \simeq \gamma \rho \left.\frac{4 \pi}{3} H^{-3} \right|_{t=t_\text{i}} \nonumber \\
 &\simeq \gamma M_\text{eq} \sqrt{2} \left( \frac{g_{*,\text{eq}}}{g_{*,\text{i}}} \right)^{1/2} \left( \frac{T_\text{eq}}{T_\text{i}} \right)^2 \nonumber \\
 & \simeq 10^4 \,  \text{g}\,  \left(\frac{\gamma}{0.2} \right) \left( \frac{g_{*,\text{i}}}{106.75} \right)^{-1/2} \left( \frac{T_\text{i}}{4.3\times 10^{13}\, \text{GeV}} \right)^{-2}, 
 \label{eq:mpbh_k}
\end{align}
or by its inversion as
\begin{align}
T_\text{i} \simeq 4.3 \times 10^{13} \,\text{GeV} \left( \frac{\gamma}{0.2} \right)^{1/2}  \left( \frac{g_{*,\text{i}}}{106.75} \right)^{-1/4}  \left( \frac{ M_{\text{PBH,i}}}{10^4 \, \text{g}} \right)^{-1/2},  \label{eq:T_i}
\end{align}
where $\rho$ represents the energy density, $H$ is the Hubble parameter, and $M_\text{eq} (\simeq 5.9\times 10^{50} \,\text{g})$ is the horizon mass at the late-time equality time ($z \sim 3400$).
Here we do not take into account the effects of the critical collapse phenomena on the PBH mass~\cite{Niemeyer:1997mt,Niemeyer:1999ak,Shibata:1999zs,Musco:2004ak} for simplicity.  (We will briefly come back to this point in the conclusion section, Sec.~\ref{sec:conclusions}.)  
$\gamma$ is the fraction of the PBH mass in the horizon mass at the formation, which is analytically estimated as $\gamma \sim (1/\sqrt{3})^3 \sim 0.2$ for the PBH production during a RD era~\cite{Carr:1975qj}. We take $\gamma = 0.2$ as a fiducial value in the following.

Finally, we discuss the initial PBH abundance at the production required to have the PBH-dominated era and the relation between the PBH abundance and the length of the PBH-dominated era. 
For this purpose, let us start with the evolution of the energy density of PBHs
\begin{align}
  \frac{\rho_{\text{PBH}}}{\rho_{\text{PBH,i}}} = \left(\frac{a_\text{i}}{a} \right)^3 = \frac{g_{s*} T^3}{g_{s*,\text{i}} T_\text{i}^3},
\end{align}
where we have used the entropy conservation law.
On the other hand, the radiation energy density can be written as
\begin{align}
  \frac{\rho_{\text{r}}}{\rho_{\text{r,i}}} = \frac{g_* T^4}{g_{*,\text{i}} T_\text{i}^4}.
\end{align}
Using these relations, we can express the initial PBH fraction as
\begin{align}
  \beta &\equiv \frac{\rho_{\text{PBH,i}}}{\rho_{\text{tot,i}}} \nonumber \\
  & \simeq \frac{\rho_{\text{PBH,i}}}{\rho_{\text{r,i}}} = \frac{g_*}{g_{*,\text{i}}} \frac{g_{s*,\text{i}}}{g_{s*}} \frac{T}{T_\text{i}} \frac{\rho_\text{PBH}}{\rho_\text{r}},
\end{align}
where $\rho_\text{tot}$ represents the total energy density.
In the second line, we have assumed that the radiation dominates the total energy density at the PBH production.
Note that the initial PBH fraction $\beta$ can also be interpreted as the PBH formation probability in a given Hubble patch.
Since we focus on the early Universe, $g_*=g_{s*}$ is satisfied.
Now one can easily see that the PBH-dominated era, $\rho_\text{PBH} > \rho_\text{r}$, can be realized when the initial PBH fraction satisfies
\begin{align}
  \beta  > \beta_{\text{min}} \equiv \frac{T_\text{R}}{T_\text{i}},
  \label{eq:beta_min_def}
\end{align}  
where $T_\text{R}/T_\text{i}$ can be evaluated from Eqs.~\eqref{eq:tr_pbh_mass} and \eqref{eq:T_i} as
\begin{align}
  \frac{T_\text{R}}{T_\text{i}} &= 6.5 \times 10^{-10} \left( \frac{T_\text{i}}{4.3 \times 10^{13}\, \text{GeV}} \right)^{-1} \left( \frac{M_{\text{PBH,i}}}{10^4 \, \text{g}} \right)^{-3/2} \nonumber \\
  & \qquad \qquad \qquad \times  \left( \frac{g_{\text{H}*}(T_\text{PBH})}{108} \right)^{1/2} \left( \frac{g_*(T_\text{R})}{106.75} \right)^{-1/4} \nonumber \\
  & = 6.5 \times 10^{-10} \left( \frac{M_{\text{PBH,i}}}{10^4 \, \text{g}} \right)^{-1}  \left( \frac{g_{\text{H}*}(T_\text{PBH})}{108} \right)^{1/2}.
\end{align}
Recall that $T_{\text{i}}$ represents the temperature of the Universe at the PBH production.

The ratio $\beta/\beta_\text{min}$ is related to the energy density at the reheating
\begin{align}
  \frac{\beta}{\beta_\text{min}} = \frac{T_{\text{eq},1}}{T_\text{R}} = \left( \frac{g_{*,\text{eva}} }{ g_{*,\text{eq},1}} \right)^{1/4} \left(\frac{ \rho_{\text{PBH,eq},1}}{ \rho_{\text{tot,eva}}}\right)^{1/4}, 
\end{align}
where the subscript ``$\text{eq},1$'' represents the value when $\rho_\text{PBH} = \rho_\text{r}$ is satisfied at the beginning of the PBH-dominated era. 
This equation implies that the length of the PBH-dominated era is related to the PBH fraction
\begin{align}
  \frac{\beta}{\beta_{\text{min}}} \simeq \left( \frac{g_{*,\text{eva}}}{ g_{*,\text{eq},1}} \right)^{1/4} \left(\sqrt{2} -1\right)^{3/2} \left(\frac{\eta_{\text{eva}}}{\eta_{\text{eq},1}} \right)^{3/2},
  \label{eq:beta_eta_rel}
\end{align}
where we have used the approximated relation~\cite{Mukhanov:991646}
\begin{align}
  \frac{a}{a_\eqf} \simeq \left(\sqrt{2}-1 \right)^2 \left( \frac{\eta}{\eta_\eqf} \right)^2 \quad \text{for}~~\eta_\eqf \ll \eta \leq \eta_\text{eva}.
\end{align}
The conformal time is given by $\eta = \int \dd t / a$.
Note that Eq.~(\ref{eq:beta_eta_rel}) is valid for $\eta_\eqf \ll \eta_\text{eva}$.

Using the abundance $\beta$, we can express the wavenumber corresponding to the PBH formation scale in terms of the PBH mass. 
Taking into account the presence of the PBH-dominated era (assuming $\beta > \beta_\text{min}$), it is given by
\begin{align}
k_\text{i} = & \left( \frac{g_{s*}(T_{\text{R}-}) g_{s*}(T_0)}{g_{s*}(T_\text{i}) g_{s*}(T_\text{R}) } \right)^{1/3} \beta_\text{min} \left( \frac{\beta}{\beta_\text{min}} \right)^{-1/3} \frac{T_0}{T_\text{R}} \frac{4 \pi \gamma M_\text{Pl}^2}{M_{\text{PBH,i}}} \nonumber \\
 =& 1.43 \times 10^{20} \, \text{Mpc}^{-1}  \left( \frac{\beta}{10^{-7}} \right)^{-1/3}  \left( \frac{M_\text{PBH,i}}{10^4 \, \text{g}} \right)^{-5/6} , \label{k_i}
\end{align}
where $T_{\text{R}-}$ ($\neq T_\text{R}$) is the temperature just before the evaporation when we regard the evaporation as a sudden event.
 In the second line, we have omitted the $g_{*}$ dependence to have a simple expression (we have assumed $g_{s,*}(T_\text{i}) = g_{s,*}(T_\text{eq,1}) =g_{s,*}(T_{\text{R}-}) = g_{s,*}(T_\text{R}) \simeq 106.75 $). 
Note that the PBH evaporation causes the entropy production and therefore $k_i$ depends on $\beta$.
Comparing the scales associated to the PBH formation (Eq.~\eqref{k_i}) and to the PBH evaporation (Eq.~\eqref{eq:kr_tr}) , we see that the GWs induced right after the evaporation have a smaller typical frequency than those produced by other mechanisms.

\section{Evolutions of scalar perturbations}
\label{sec:evo_pertb}

Since the enhancement of GWs strongly depends on the oscillation amplitude of the scalar perturbations after the reheating transition (i.e.~evaporation), we discuss the evolutions of the perturbations around the transition here. 
The results in this section are used in the next section to calculate the induced GWs.

\subsection{Formulas and numerical results}

In this subsection, we introduce the formulas to calculate the perturbations and show the numerical results.
The metric perturbations in the conformal Newtonian gauge can be written as
\begin{align}
  \dd s^2 = &a^2 \left[ -\left(1+2\Phi \right) \dd \eta^2 \phantom{\frac{1}{2}} \right. \nonumber \\
  & \qquad \qquad \left. + \left( \left(1-2\Psi \right) \delta_{ij} + \frac{1}{2} h_{ij} \right) \dd x^i \dd x^j \right],
\end{align}
where $h_{ij}$ is the tensor perturbation, which satisfies $h^i_i = 0$ and $\partial h_{ij}/\partial x_i = 0$.
In the synchronous gauge, they can be written as
\begin{align}
  \dd s^2 = a^2 \left[ -\dd \eta^2 + \left(\delta_{ij} + H_{ij} + \frac{1}{2} h_{ij} \right) \dd x^i \dd x^j \right],
\end{align}
where, in Fourier space, $H_{ij} = \hat k_i \hat k_j \gamma + (\hat k_i \hat k_j - \frac{1}{3} \delta_{ij}) 6 \epsilon$.\footnote{$\gamma$ and $\epsilon$ correspond to $h$ and $\eta$ in Ref.~\cite{Ma:1995ey}, respectively.}

As we will introduce in the next section, we use the formulas for the induced GWs which are derived in the conformal Newtonian gauge. 
In addition, since we focus on the early Universe, we can assume that there is no anisotropic stress and take $\Psi = \Phi$.
Therefore, all we need to understand is the evolution of the gravitational potential $\Phi$ around the transition.
In this paper, to make discussion easier, we first calculate the perturbations in the synchronous gauge, and then, we transform them to quantities in the Newtonian gauge.
This is because, in the synchronous gauge, the decay rate (PBH evaporation rate) does not depend on the spatial coordinates, and therefore the situation is simpler than that in the Newtonian gauge.
The situation of the PBH evaporation is similar to the decaying dark matter scenario, and we express the PBH quantities as non-relativistic matter quantities, e.g.~$\rho_\PBH \rightarrow \rho_\text{m}$ to match the convention used in Ref.~\cite{Poulin:2016nat}.

First, we discuss background quantities. 
The Friedmann equation reads
\begin{align}
  \mathcal H = \frac{a}{\sqrt{3}M_\text{Pl}} \sqrt{\rho_\text{tot}}, 
\end{align}
where the comoving Hubble parameter is defined by $\mathcal H = a'/a$ with $a'=\dd a/ \dd \eta$. 
The derivatives of background quantities are given by~\cite{Poulin:2016nat}
\begin{align} 
  \label{eq:rho_m_evo}
  \rho'_\text{m} &= - \left( 3\mathcal H - \frac{\dd \, \text{ln}\, M_\text{PBH}}{\dd \eta} \right) \rho_\text{m}, \\
  \rho'_\text{r} &= -4\mathcal H \rho_\text{r} -\frac{\dd \, \text{ln}\, M_\text{PBH}}{\dd \eta} \rho_\text{m}, 
\end{align}
where $\rho_\text{m}$ and $\rho_\text{r}$ are the energy densities of the non-relativistic matter (PBHs) and radiation. 
Note here that $-\dd \ln M_{\text{PBH}} / \dd \eta$ is the PBH decay rate per conformal time $a \Gamma$ where $\Gamma$ via evaporation is given in Eq.~(\ref{eq:decay_rate}).

Next, we discuss the perturbations in the synchronous gauge.
Here, we introduce the perturbation as $\delta = \delta \rho / \rho$ where $\delta \rho$ is the perturbed energy density.
The fluctuation $\delta \rho_\text{m}$ originates from the fluctuation of the number density of PBHs. 
We also introduce the velocity divergence $\theta \equiv \partial v_i/\partial x_i$ where $v_i$ is the fluid velocity, and take the coordinates that always satisfy $\theta_\text{m}=0$~\cite{Ma:1995ey}. 
Then, we obtain the following equation~\cite{Poulin:2016nat}:
\begin{align}
  \delta'_\text{m} &= - \frac{\gamma'}{2}. 
\end{align}
For radiation perturbations, we get
\begin{align}
  \delta'_\text{r} &= -\frac{4}{3} \left(\theta_\text{r} + \gamma'/2 \right) - \frac{\dd\, \text{ln} \, M_\text{PBH}}{\dd \eta} \frac{\rho_\text{m}}{\rho_\text{r}} \left(\delta_\text{m} - \delta_\text{r} \right), \\
  \theta'_\text{r} &= \frac{k^2}{4} \delta_\text{r} + \frac{\dd\, \text{ln} \, M_\text{PBH}}{\dd \eta} \frac{\rho_\text{m}}{\rho_\text{r}} \theta_\text{r} ,
\end{align}
where we have neglected the anisotropic stress. 
The equations of motion for the metric perturbations are given by~\cite{Ma:1995ey}
\begin{align}
  k^2 \epsilon - \frac{1}{2} \frac{a'}{a} \gamma' 
  &= -\frac{3}{2} \mathcal H^2 \left( \frac{\rho_\text{m}}{\rho_\text{tot} } \delta_\text{m} + \frac{\rho_\text{r}}{\rho_\text{tot}} \delta_\text{r} \right), \\
  k^2 \epsilon' 
  & = 2 \mathcal H^2 \frac{\rho_\text{r}}{\rho_\text{tot}} \theta_\text{r}.
\end{align}
Following Ref.~\cite{Ma:1995ey}, we take the initial conditions of the perturbations as follows:
\begin{align}
  &\delta_\text{r} = - \frac{2}{3}C\left(k \eta\right)^2 ,\  \delta_\text{m} = \frac{3}{4} \delta_\text{r},
  \  \theta_\text{r} = - \frac{1}{18}C \left(k^4 \eta^3 \right), \nonumber\\
  \label{eq:sync_initial}
  & \gamma = C\left(k\eta\right)^2,\  \epsilon = 2C - \frac{1}{18} C\left(k\eta\right)^2,
\end{align}
where the coefficient $C$ is related to the curvature perturbations as $C = \zeta/2$ on superhorizon scales.
Since the initial conditions are derived by assuming a RD era in Ref.~\cite{Ma:1995ey}, we start the numerical calculation much before the PBH-dominated era starts.

\begin{figure}[tbh] 
        \centering \includegraphics[width=1\columnwidth]{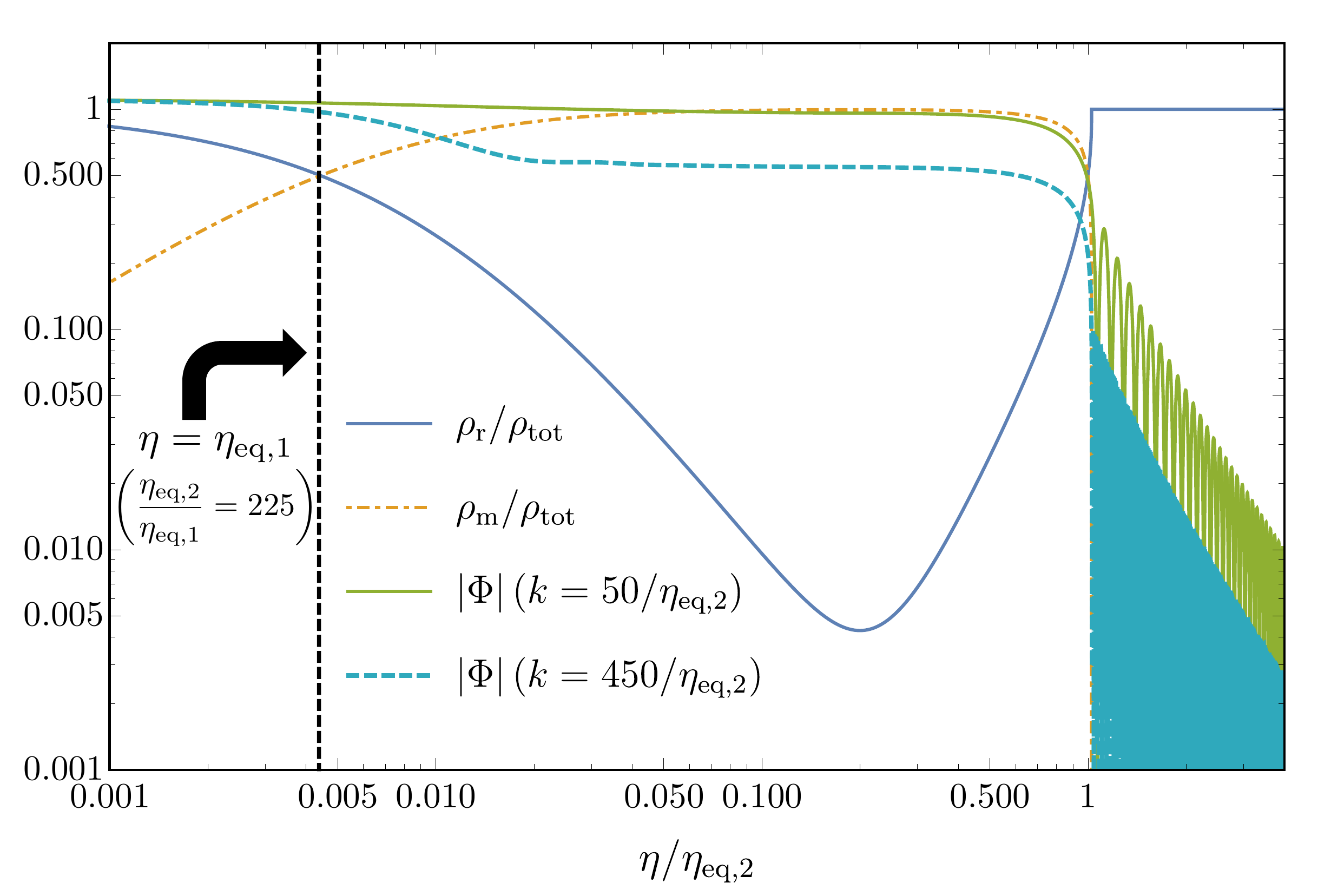}
        \caption{
    Evolutions of the background energy densities and the transfer function of the gravitational potential, $\Phi$. 
        }
        \label{fig:back_phi_evo}
\end{figure}

$\Psi$ in the conformal Newtonian gauge is related to $\epsilon$ and $\gamma$ in the synchronous gauge as
\begin{align}
  \label{eq:sync_to_new}
  \Psi &= \epsilon - \mathcal H \alpha,
\end{align}
with $\alpha \equiv (6 \epsilon + \gamma)'/(2k^2)$. 
Note again that we can safely take $\Psi = \Phi$ in the regime of our interest.
We numerically calculate the perturbations around the transition which are governed by Eqs.~(\ref{eq:rho_m_evo})-(\ref{eq:sync_initial}).
Then, by using the transformation in Eq.~(\ref{eq:sync_to_new}), we acquire the time evolution of the gravitational potential $\Phi$ in the Newtonian gauge.

Figure~\ref{fig:back_phi_evo} shows the evolutions of the energy densities and the transfer function of the gravitational potential.\footnote{
	Evolutions of other background quantities, such as the scale factor and the entropy density, are discussed in Ref.~\cite{Gutierrez:2017ibk}.
}
The transfer function is defined as the gravitational potential that is normalized as $\Phi = 10/9$ on the superhorizon scales during the eRD era that precedes the PBH-dominated era.
This normalization corresponds to that taken in Ref.~\cite{Inomata:2019ivs}.
Here, we define $\eta_{\text{eq},2}$ as the equality time, satisfying $\rho_\text{m} = \rho_\text{r}$, around the reheating ($\eta_{\text{eq},1} < \eta_{\text{eq},2}$).
The difference between $\eta_{\text{eq},2}$ and $\eta_\text{eva}$ is very small in Fig.~\ref{fig:back_phi_evo} and therefore we can use the two conformal times interchangeably when we estimate the order of magnitude of the induced GWs, which is one of the goals of this paper.
From this figure, we can also see that the gravitational potential slightly decays at the reheating ($\sim \eta_\eqs$) and starts to oscillate with the amplitude of $\Phi \sim \mathcal O(0.1)$ after the reheating. 
Note that, if the transition is exactly sudden, the gravitational potential does not decay during the transition~\cite{Inomata:2019ivs}.
In this sense, the reheating is not completely sudden, but more sudden than that in the case with a constant decay rate, discussed in Ref.~\cite{Inomata:2019zqy}.

We can also see that the gravitational potential for a small scale $k=450/\eta_{\text{eq},2}$ is less than unity even before $\eta_{\text{eq},2}$.
This is because this perturbation reenters the horizon so early that the Universe has not yet been completely dominated by PBHs.
Therefore, the perturbation decays a little bit until PBHs dominate the Universe.

\subsection{Wavenumber dependence of the suppression of gravitational potential} 

The enhancement of the induced GWs is caused by the fast oscillations of $\Phi$ after the transition~\cite{Inomata:2019ivs}.
Since the power spectrum of the GWs depends on the fourth power of $\Phi$, it is essential to estimate its oscillation amplitude precisely. 
For this purpose, we define the normalization factor $S(\leq 1)$ as the amplitude of $\Phi$ when it starts to oscillate around $\eta_\text{eva}$ as in Fig.~\ref{fig:back_phi_evo} (see also Fig.~\ref{fig:phi_fit} below).
Namely, this factor $S$ characterizes how sudden the transition is. Roughly speaking, $S$ is close to unity when the perturbation enters the horizon during the completely PBH-dominated era and the reheating transition is sudden, while it becomes small when the perturbation enters the horizon before the PBH-dominated era or when the reheating transition is gradual. 
In the first half of this subsection, we derive an analytic approximation formula for the normalization factor.  In the latter half, it is numerically computed, and they are compared with each other.  

Let us start with a discussion on the wavenumber dependence of the normalization factor $S$.
First, we focus on the suppression occurring soon after the horizon entry.
A corresponding example in Fig.~\ref{fig:back_phi_evo} is shown in the cyan dashed line, i.e.~$\Phi$ with $k=450/\eta_\eqs$ around $\eta/\eta_\eqs \simeq 0.01$.
This suppression occurs because of the remaining energy density of radiation around $\eta \sim \eta_\eqf$.
After the energy density of PBHs becomes much larger than that of radiation, the gravitational potential becomes constant (the plateau in Fig.~\ref{fig:back_phi_evo}).
The evolution of the perturbations during the transition from the eRD era to the PBH-dominated era is the same as that during the transition from the late RD era to the late MD era at $z \sim 3400$. 
The wavenumber dependence of the constant value of the gravitational potential during the PBH-dominated era, whose transfer function is dubbed $\Phi_\text{plateau}$, can be fitted by the following function~\cite{Bardeen:1985tr,Dodelson:1282338}:
\begin{widetext}
\begin{align}
  \Phi_\text{plateau} (x_\eqf) &\equiv \Phi (x)|_{\eta_{\text{eq,1}} \ll \eta \lesssim \eta_{\text{eq},2}} \nonumber \\
  &\simeq \frac{\text{ln}[1+0.146\, x_\eqf]}{\left(0.146\, x_\eqf \right)}\left[ 1 + 0.242\, x_\eqf + \left(1.01\, x_\eqf \right)^2 
+ \left(0.341\, x_\eqf \right)^3 + \left(0.418\, x_\eqf \right)^4 \right]^{-0.25},    \label{eq:Phi_plateau}
\end{align}
\end{widetext}
where $x_\eqf \equiv k \eta_\eqf$ and $\Phi_\text{plateau}$ is normalized as $\Phi_\text{plateau}(x_\eqf \rightarrow 0) \rightarrow 1$.

Next, we discuss the decay of the gravitational potential during the reheating.
According to Ref.~\cite{Inomata:2019zqy}, the decay of $\Phi$ during the transition can be approximated as 
\begin{align}
  \frac{\Phi(t)}{\Phi_\text{plateau}} \simeq& \exp\left( - \int^t_{t_\text{i}} \dd \bar t \, \Gamma (\bar t)\right) \nonumber \\
  = & \frac{\left(3 t_\text{eva} - 3t\right)^{1/3}}{\left(3t_\text{eva} - 3 t_\text{i}\right)^{1/3}} \nonumber \\
  \simeq & \left( 1 - \left(\frac{t}{t_\text{eva}} \right)\right)^{1/3},
  \label{eq:phi_decay_ana}
\end{align}
where $t_\text{i} (\ll t_\text{eva})$ is the time at the PBH formation, and we have used Eq.~(\ref{eq:decay_rate}).
After a while, $\Phi$ stops to follow Eq.~(\ref{eq:phi_decay_ana}) as shown in Fig.~\ref{fig:back_phi_evo} since $\Phi$ decouples from the matter perturbation.  Then, it starts to oscillate with its amplitude decaying relatively slowly ($\sim a^{-2}$). 
Since Eq.~(\ref{eq:phi_decay_ana}) is derived with the assumption $|\ddot \Phi| \ll k^2/(3a^2) |\Phi|$ as a necessary condition, we expect that the decoupling occurs when or before the inequality becomes invalid.
This means 
\begin{align}
  \left| \frac{\ddot \Phi}{\Phi} \right|_{t=t_\text{dec}} \simeq  \frac{2}{9\left(t_\text{dec} - t_\text{eva}\right)^2} \lesssim \frac{k^2}{3a^2},
  \label{eq:dec_cond}
\end{align}
where the dot represents a derivative with respect to $t$ and $t_\text{dec}$ is the decoupling time.
Then, we can define the lower bound of the normalization factor $S(k)$ as 
\begin{align}
  S_\text{low}(k) &\equiv  \left( 1 - \left( \frac{t_\text{dec}}{t_\text{eva}} \right) \right)^{1/3} \Phi_\text{plateau}(x_\eqf) \nonumber \\
&\simeq \left( \frac{\sqrt{6}}{k \eta_\text{eva} } \right)^{1/3} \Phi_\text{plateau}(x_\eqf) ,  \label{eq:s_low} 
\end{align}
where we have used the relation $\eta a = 3t$, valid during the PBH-dominated era. 

In the following, we compare the above analytic estimations with numerical calculations. 
In the RD era after the reheating due to PBH evaporation, the evolution of the gravitational potential is given as the solution of the following equation~\cite{Mukhanov:991646}:
\begin{align}
\Phi'' +  4 \mathcal H \Phi' +  \frac{k^2}{3} \Phi = 0.
\label{eq:phi_evo_eq}
\end{align} 
To quantify the decay during the transition, we define the fitting formula for $\Phi$ as
\begin{align}
  \Phi_\text{osc,fit} (x,x_0) =  S \left(A(x_0) \mathcal J(x,x_0) + B(x_0) \mathcal Y(x,x_0) \right),
  \label{eq:phi_osc_fit}
\end{align}
with $x = k\eta$.
Here $S$ and $x_0$ are fitting parameters, which describe the suppression of $\Phi$ before its oscillation and the start time of the oscillation, respectively.
$\mathcal J(x)$ and $\mathcal Y(x)$ are independent solutions for Eq.~(\ref{eq:phi_evo_eq}), which can be written with the first and second spherical Bessel functions, $j_1(x)$ and $y_1(x)$, as
\begin{align}
\mathcal J(x,x_0) = \frac{ 3\sqrt{3} \,  j_1\left(\frac{x-x_0/2}{\sqrt{3}} \right)}{x-x_0/2}, \ 
\mathcal Y(x,x_0) = \frac{3\sqrt{3} \, y_1\left( \frac{x - x_0/2}{\sqrt{3}} \right)}{x-x_0/2}.
\label{eq:c_d_formula}
\end{align}
We determine the coefficients $A(x_0)$ and $B(x_0)$ so that $\Phi(x_0) = S$ and $\Phi'(x_0) = 0$:
\begin{align}
\label{eq:a_formula}
A(x_0) &= \frac{1}{\mathcal J(x_0) - \frac{\mathcal Y(x_0)}{\mathcal Y'(x_0)} \mathcal J'(x_0)}, \\
\label{eq:b_formula}
B(x_0) &= - \frac{\mathcal J'(x_0)}{\mathcal Y'(x_0)} A(x_0).
\end{align}
Note that, if the transition is exactly sudden as discussed in Ref.~\cite{Inomata:2019ivs}, the approximation formula with $x_0 = x_\R$ and $S=\Phi_\text{plateau}$ fits the numerical result, where $x_\R = k\eta_\text{R}$ and $\eta_\text{R}$ is the conformal time at the sudden-limit reheating.
From this, we expect that $x_0 \simeq k \eta_\text{eva}$ fits the numerical results well in the situation we consider here.
In fact, by numerically finding the optimal value of $x_0$, we confirm $x_0 \simeq k \eta_\text{eva}$.
Figure~\ref{fig:phi_fit} shows the numerical result and the approximation formula with the fitted parameters.
We can see that the approximation formula agrees with the oscillation part of the numerical result.

\begin{figure}[htb] 
        \centering \includegraphics[width=1\columnwidth]{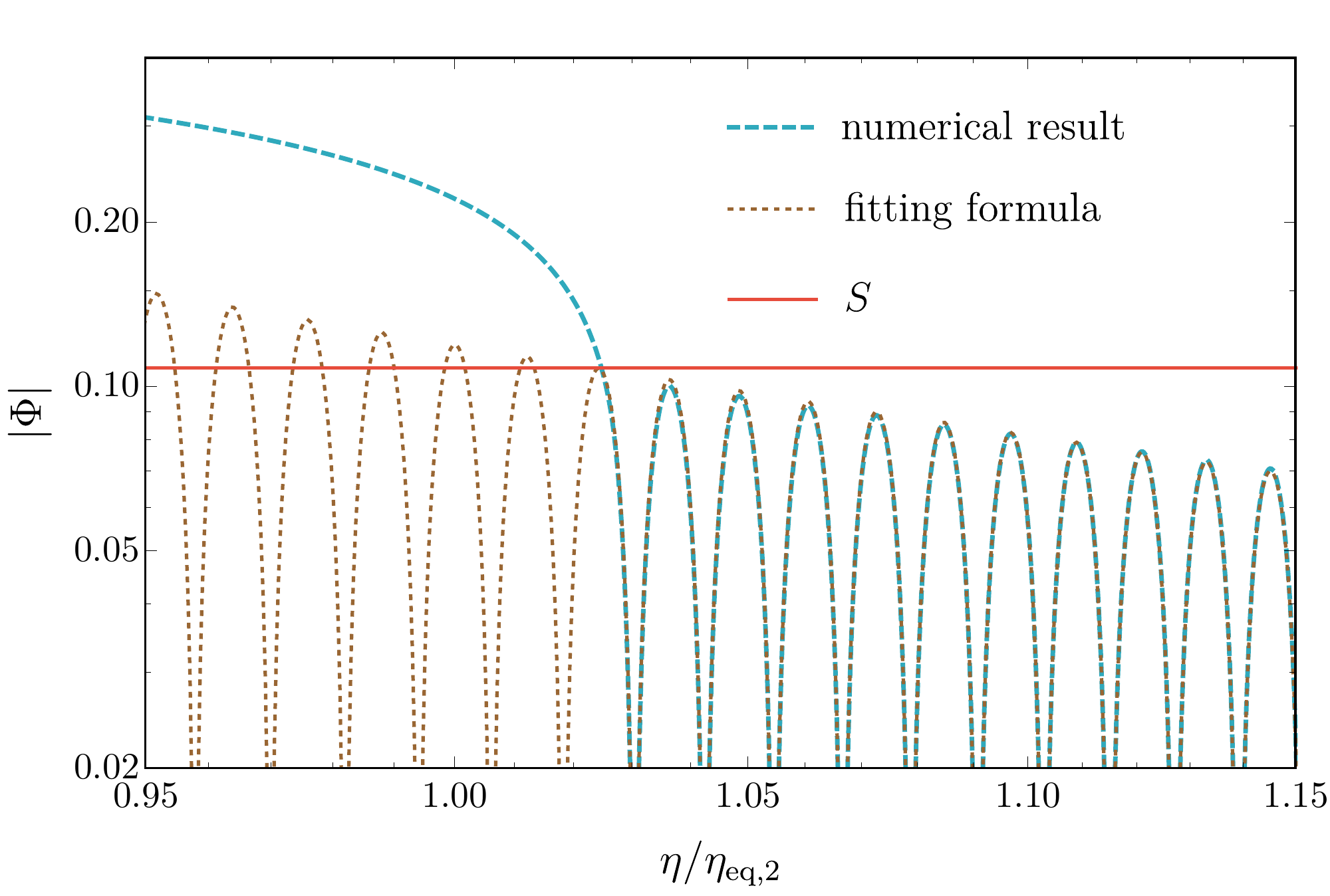}
        \caption{
        The evolution of $\Phi$ with $k=450/\eta_\eqs$, stretched around the transition.
        The numerical result and the fitting formula for the oscillation, given in Eq.~(\ref{eq:phi_osc_fit}), are plotted with a cyan dashed and a brown dotted line, respectively.
        $S = 0.108$ and $x_0 = 236$ are taken as the fitted parameters.
        The normalization factor $S (=0.108)$ is also plotted with a red solid line.
        }
        \label{fig:phi_fit}
\end{figure}

\begin{figure}[htb] 
  \centering \includegraphics[width=1\columnwidth]{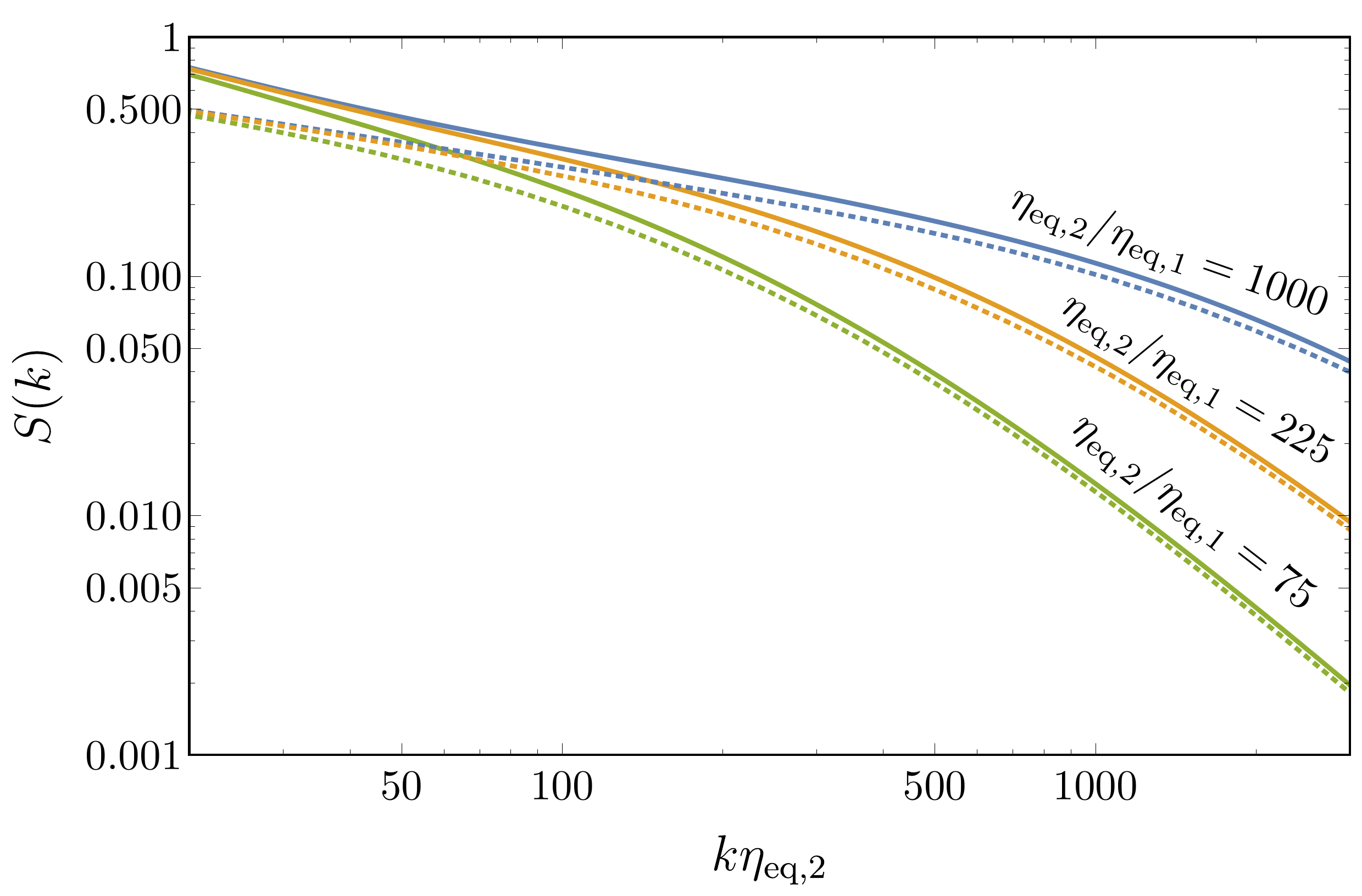}
        \caption{
        Wavenumber dependence of the normalization factor with $\eta_{\text{eq},2}/\eta_{\text{eq},1} = 1000$ (blue, top), $225$ (orange, middle), and $75$ (green, bottom). 
        Solid lines show the numerical results, and dotted lines show the lower bounds of the normalization factor $S_\text{low}$ (see Eq.~\eqref{eq:s_low}). 
        }
        \label{fig:s_low}
\end{figure}

Figure~\ref{fig:s_low} shows the wavenumber dependence of the normalization factor with different lengths of the PBH-dominated era, characterized by $\eta_\eqs/\eta_\eqf = 1000$, $225$, and $75$.  The lower bounds (dashed lines; Eq.~\eqref{eq:s_low}) and the numerical results (solid lines; Eqs.~\eqref{eq:phi_osc_fit}-\eqref{eq:b_formula}) are compared. 
We can see that the normalization factor is close to its lower bound for $k \gtrsim 2/\eta_\eqf$.
The difference between the numerical result and the lower bound is less than $20\%$ at $k = 2/\eta_\eqf$ for $75 < \eta_\eqs/\eta_\eqf < 1000$.
It becomes smaller for a larger $k$.

\section{Gravitational waves induced by scalar perturbations}
\label{sec:induced_gws}

In this section, we discuss the GWs induced by the scalar perturbations that experience the PBH-dominated era. 
As a byproduct, we develop a method to estimate the induced GWs in the presence of the sudden reheating from an eMD era with a finite duration, which is  preceded by an eRD era.  Such multiple changes of the equation-of-state for the Universe is naturally realized in the case of a PBH-dominated era, but one can consider other cases such as moduli domination.  Some of the techniques and results in this and the subsequent sections, including the estimate on how sudden the transition should be for significant enhancement of GWs, are also applicable to such general cases.

In the following, we consider the induced GWs in the conformal Newtonian gauge and assume that the curvature perturbations follow the Gaussian distribution for simplicity.\footnote{The gauge (in)dependence of the induced GWs is discussed in Refs.~\cite{Arroja:2009sh,Hwang:2017oxa,Gong:2019mui,Tomikawa:2019tvi,DeLuca:2019ufz,Inomata:2019yww,Yuan:2019fwv} and the effects of the non-Gaussianity are discussed in Refs.~\cite{Nakama:2016gzw,Garcia-Bellido:2017aan,Cai:2018dig,Unal:2018yaa}.}

\subsection{Basic formulas}

Here, we briefly summarize the formulas for the induced GWs which are introduced in Ref.~\cite{Inomata:2019ivs}, though some of them are modified to fit the situation of the reheating by the PBH evaporation.
For the moment, we regard the PBH-dominated era suddenly ends at $\eta_\text{eva} (\simeq \eta_{\text{eq},2})$ for simplicity (see Fig.~\ref{fig:back_phi_evo}).  We will explain how to take into account the fact that it is not completely sudden later (below Eq.~\eqref{eq:green_rd}). 

Since the scale factor and the Hubble parameter are continuous at $\eta_\text{eva}$, their time dependences can be expressed as~\cite{Mukhanov:991646}
\begin{align}
\label{eq:a_express}
\frac{a(\eta)}{a(\eta_{{\text{eq},1}})} =& \begin{cases}
\left( \cfrac{\eta}{\eta_*} \right)^2 + 2 \left( \cfrac{\eta}{\eta_*} \right) & (\eta \leq \eta_\text{eva}) \\[10pt]
\cfrac{ 2 \eta \left(\eta_\text{eva} + \eta_*\right) - \eta_\text{eva}^2}{\eta_*^2}  & (\eta > \eta_\text{eva})
\end{cases}, \\[5pt]
\mathcal{H}(\eta)=& \begin{cases}
  \cfrac{2\eta + 2 \eta_*}{\eta^2 + 2 \eta \eta_*} & (\eta \leq \eta_\text{eva}) \\[10pt]
  \cfrac{1}{\eta - \frac{\eta_\text{eva}^2}{2\left(\eta_\text{eva} + \eta_*\right)} } & (\eta > \eta_\text{eva})
\end{cases},
\end{align}
where $\eta_* = \eta_\eqf/\left(\sqrt{2}-1\right)$.
The energy density parameter of the induced GWs per logarithmic interval in $k$ is given by
\begin{align}
\Omega_{\rm{GW}} (\eta, k) &= \frac{\rho_{\rm{GW}} (\eta, k) } { \rho_{\rm{tot}}(\eta)} \nonumber\\ 
&=
\frac{1}{24} \left( \frac{k}{a(\eta) H(\eta) } \right)^2 
\overline{\mathcal P_{h} (\eta,k)} ,
\label{eq:gw_formula}
\end{align}
where $\overline{\mathcal P_h(\eta, k)}$ is the time-averaged power spectrum of GWs, which is related to the power spectrum of the curvature perturbations as~\cite{Inomata:2016rbd,Kohri:2018awv}
\begin{align}
\overline{\mathcal P_{h} (\eta,k)} = 4\int^\infty_0 \dd v &\int^{1+v}_{|1-v|} \dd u \left[ \frac{4v^2 - (1+v^2 - u^2)^2}{4vu}  \right]^2 \nonumber \\
& \times \overline{I^2(u,v,k,\eta,\eta_\text{eva}) }\mathcal P_\zeta(u k) \mathcal P_\zeta(v k).
\label{eq:p_h_formula}
\end{align}
$I(u,v,k,\eta,\eta_\text{eva})$ describes the evolution of the scalar perturbation and can be expressed as
\begin{align}
I(u,v,k,\eta,\eta_\text{eva}) =& \int^{x}_0 \dd \bar{x} \frac{a(\bar \eta)}{a(\eta)} k G_k(\eta, \bar \eta) f(u,v,\bar x, x_\text{eva}),
\label{eq:i_formula}
\end{align}
where $x_{\text{eva}} = k \eta_{\text{eva}}$.
$G_k(\eta,\bar \eta)$ is the Green function satisfying the following equation:
\begin{align}
G_k''(\eta, \bar \eta) + \left( k^2 - \frac{a''(\eta)}{a(\eta)} \right) G_k(\eta, \bar \eta) = \delta (\eta - \bar \eta),
\label{eq:g_formula}
\end{align}
where the prime denotes the derivative with respect to $\eta$, not $\bar \eta$.
The function $f(u,v,\bar x, x_{\text{eva}})$ in Eq.~(\ref{eq:i_formula}) is expressed with the transfer functions of the gravitational potential ($\Phi$) as 
\begin{align}
f(u,v,\bar{x}, x_\text{eva})= \frac{3}{25(1+w)} &\left[ 2(5+3w) \Phi(u\bar{x})\Phi(v\bar{x}) \phantom{\mathcal H^{-2}} \right. \nonumber \\
& +4 \mathcal H^{-1} \left(\Phi'(u\bar{x})\Phi(v\bar{x}) + \Phi(u\bar{x})\Phi'(v\bar{x})\right) \nonumber \\
&\left. + 4 \mathcal H^{-2} \Phi'(u\bar{x})\Phi'(v\bar{x}) \right]
\label{eq:f_def}
\end{align}
where $w$ is the equation-of-state parameter, defined as $w \equiv P/\rho$ with $P$ being the pressure.
We take the same normalization of $\Phi$ as in Fig.~\ref{fig:back_phi_evo} ($\Phi(x \rightarrow 0) = 10/9$).
Note that $\Phi(x)$ also depends on $x_\text{eva}$ implicitly, so, e.g., $\Phi(u \bar{x})$ actually means $\Phi (u\bar{x}, ux_{\text{eva}})$.

The evolution of the transfer function is discussed in Sec.~\ref{sec:evo_pertb}.
If there is a PBH-dominated era in the early Universe, the dominant contribution comes from the fast oscillations of $\Phi$ at $\eta > \eta_\text{eva}$~\cite{Inomata:2019ivs}.
For the perturbations entering the horizon much before the reheating where $k\eta_\text{eva} \gg 1$, the last term in Eq.~(\ref{eq:f_def}) 
dominates because it behaves as $\mathcal H^{-2} (\Phi')^2 \sim (k \eta_\text{eva})^2 \Phi^2$ soon after the reheating.
Then, the enhanced source term leads to the amplification of the induced GWs. 
This is the Poltergeist mechanism (see also Ref.~\cite{Inomata:2019ivs} for a detailed explanation of the enhancement of the induced GWs). 

A more physical explanation of the Poltergeist mechanism is given as follows.
During the PBH-dominated era, the density perturbations grow proportionally to the scale factor, and the scalar source term for each $k$ is kept constant even for the subhorizon modes.  The density perturbations are nothing but the PBH number-density fluctuations, so they do not oscillate.  After the reheating by the PBH evaporation, PBHs and their fluctuations are converted to radiation and its fluctuation.  The fluctuation of radiation is nothing but the sound waves of the thermal bath.  These sound waves oscillate with their enhanced amplitudes because the density perturbations have grown until the evaporation and they do not have enough time to decay because of the sudden transition.  
When the sources oscillate, there is generally a possibility of resonance.  In fact, the dominant contribution to the induced GWs on small scales 
 comes from the resonant production. (The resonance condition is also explained in Appendix~\ref{app:app_formula}.)  Note that the resonance can only happen in the RD era simply because the density perturbations, as well as the gravitational potential, do not oscillate during the eMD era.  
 This clearly highlights the fact that the GW production by the Poltergeist mechanism occurs after the PBH evaporation in contrast, e.g., to the GWs emission by Hawking radiation. There are nonzero contributions from the eMD era~\cite{Assadullahi:2009nf, Baumann:2007zm, Inomata:2019zqy}, but they are subdominant in the sudden transition case~\cite{Inomata:2019ivs}. 
 
From these observations, we neglect the contribution during $\eta<\eta_\text{eva}$ in the following and approximate the function $I(u,v,k,\eta,\eta_\text{eva})$, defined in Eq.~(\ref{eq:i_formula}), as~\cite{Inomata:2019ivs}
\begin{align}
I(u,v,k,\eta,\eta_\text{eva}) \simeq& \int^{x}_{x_\text{eva}} \dd \bar{x} \left(\frac{2\left(\bar x/x_\text{eva}\right)-1}{2\left(x/x_\text{eva}\right)-1} \right) \nonumber \\
& \qquad \times k G^\text{RD}_k(\eta, \bar \eta) f(u,v,\bar x, x_\text{eva}),
\label{eq:i_formula_RD}
\end{align}
where the Green function during the RD era is given as 
\begin{align}
  k G^\text{RD}_k (\eta, \bar \eta) = \sin(x - \bar x).
\label{eq:green_rd}
\end{align}

The enhancement of the induced GWs for the exactly sudden reheating is discussed in Ref.~\cite{Inomata:2019ivs}.
The main difference between the exactly sudden reheating scenario and the almost sudden reheating scenario, caused by PBH evaporation, lies in the normalization factor $S$.
In the case of the exactly sudden reheating, there is no suppression during the reheating, and therefore the wavenumber dependence of the normalization factor only comes from $\Phi_\text{plateau}$ as $S=\Phi_\text{plateau}$.
On the other hand, in the case of the reheating caused by the PBH evaporation, the wavenumber dependence also comes from the suppression during the reheating transition, as shown in Sec.~\ref{sec:evo_pertb} (see Fig.~\ref{fig:s_low}). 
It is not important whether the wavenumber dependence of the gravitational potential originates from the primordial curvature perturbations or from the dynamics related to the evaporation
 because the Poltergeist mechanism takes place after the PBH evaporation. 
Therefore, we can easily take into account the wavenumber dependence of $S$ by modifying the power spectrum in Eq.~(\ref{eq:p_h_formula}) as $\mathcal P_\zeta(k) \rightarrow  S^2(k)\mathcal P_\zeta(k)$ and the transfer function in Eq.~(\ref{eq:f_def}) as $\Phi \rightarrow \Phi_\text{norm}$, where 
\begin{align}
  \Phi_\text{norm}(x,x_\text{eva}) = A(x_\text{eva}) \mathcal J(x,x_\text{eva}) + B(x_\text{eva}) \mathcal Y(x,x_\text{eva}). \label{eq:Phi_norm} 
\end{align}
Although this expression describes the transfer function only for $\eta>\eta_\text{eva}$, it is sufficient to calculate the main contribution to the induced GWs, which comes from the fast oscillations of $\Phi$ after the reheating.
Note that when we perform the numerical calculation, we use $S_\text{low}(k)$ on the scale satisfying $S_\text{low}(k) < 1$ instead of the exact value of $S(k)$ for simplicity.
This is a good approximation because the main source of the GW enhancement is the scalar modes with the shortest wavelengths in the problem ($k_\text{cut} \gtrsim 2/\eta_\eqf$ where the cutoff scale $k_\text{cut}$ is introduced in the next subsection)~\cite{Inomata:2019ivs}, and $S_\text{low}(k)$ for such wavelengths is close to the exact value as shown in Fig.~\ref{fig:s_low}.

The production of the GWs due to the fast oscillations of $\Phi$ becomes inefficient after a while because the oscillation amplitude decays proportionally to $a^{-2}$.
We define the conformal time when the induced GWs become constant as $\eta_\text{c}$, where $\eta_\text{c}$ is $\mathcal O(\eta_{\text{eva}})$, much earlier than the late-time equality time.
Once the induced GWs are produced, their energy density behaves as the radiation energy density as $\rho_\text{GW} \propto a^{-4}$.
Therefore, if we take into account the suppression of the energy density parameter of the induced GWs due to the late-time evolution ($z \lesssim 3400$), the energy density parameter at the present time is given as~\cite{Ando:2018qdb}
\begin{align}
  \Omega_\text{GW}(\eta_0,k)h^2 = 0.39 \left( \frac{g_{*,\text{c}}}{106.75} \right)^{-1/3} \Omega_{\text{r},0}h^2 \Omega_\text{GW}(\eta_\text{c},k), \label{eq:current_Omega_GW}
\end{align}
where the subscript ``c'' means the value at $\eta_\text{c}$ and $\Omega_{\text{r},0}h^2 \simeq 4.2\times 10^{-5}$ is the current energy density parameter of radiation.\footnote{Throughout this paper, we neglect the masses of neutrinos for simplicity.}

\subsection{Amount of the induced GWs} 

In this subsection, we show how large the enhancement of the induced GWs is.
We assume the following primordial power spectrum for the curvature perturbations:
\begin{align} 
  \mathcal P_\zeta (k) = A_\text{s} \Theta(k_\text{cut}-k) \left( \frac{k}{k_*}\right)^{n_\text{s}-1},
  \label{eq:pzeta_almost_sc_inv}
\end{align}
where $A_\text{s}$ is the normalization, $n_\text{s}$ is the tilt, $k_*$ is the pivot scale, and  $k_\text{cut}$ is the cutoff scale.
To derive a conservative result, we take the non-linear scale $k_\text{NL}$, on which the matter perturbations $\delta_\text{m}$ becomes unity at the reheating, as $k_\text{cut}$.\footnote{
	A more fundamental cutoff scale would be the horizon scale corresponding to the time slightly after the PBH formation because one cannot neglect the discreteness of PBHs at that scale.
	If the eMD era is too short, $\delta_\text{m} = 1$ is never achieved within this fundamental cutoff scale.
}
We explain how to derive the non-linear wavenumber in Appendix~\ref{app:nl_scale}.
We leave the study about the GWs induced by the non-linear perturbations with $k>k_\text{NL}$ for future works.\footnote{There are some works discussing the GWs induced by the non-linear scalar perturbations with some uncertainties~\cite{Jedamzik:2010dq,Jedamzik:2010hq}.}
Since the normalization factor $S$ has the wavenumber dependence as $\sim k^{-7/3}$ for $k \gtrsim 2/\eta_\eqf$ up to the logarithmic factor and the dominant contribution comes from the small scale $2/\eta_\eqf \lesssim k \lesssim k_\text{NL}$, the effective tilt of the power spectrum $S^2 \mathcal P_\zeta$ can be approximated as $\sim k^{n_\text{s} - 1 - 14/3}$.
Using this value of the effective tilt and the formulas given in the appendix of Ref.~\cite{Inomata:2019ivs}, we can analytically estimate the value of $\Omega_\text{GW}$ around the resonance peak as 
\begin{align}
  \label{eq:omega_gw_res_ana}
  \Omega_\text{GW}^{(\text{res})} (\eta_\text{c},k) \simeq& 1.6 \times 10^{-6} \left(k\eta_\text{eva}\right)^7  A_\text{s}^2  \left( \frac{k}{k_*}\right)^{2 (n_\text{s}-1)}  S^4_\text{low}(k) \nonumber \\
  &\times \frac{s_0(k/k_\text{NL})}{8} \left[15 - 10 s_0^2(k/k_\text{NL}) + 3 s_0^4(k/k_\text{NL}) \right],
\end{align}
where 
\begin{align}
  s_0(k/k_\text{NL}) = \begin{cases} 
  1  & \left( k/k_\text{NL} \leq \frac{2}{1+\sqrt{3}} \right) \\
  2 \frac{k_\text{NL}}{k} - \sqrt{3}  & \left( \frac{2}{1+\sqrt{3}} \leq k/k_\text{NL} \leq \frac{2}{\sqrt{3}} \right) \\ 
  0 & \left( \frac{2}{\sqrt{3}} \leq k/k_\text{NL} \right)
  \end{cases}.
  \label{eq:s0_def}
\end{align}
See Appendix~\ref{app:app_formula} for the derivation of Eq.~(\ref{eq:omega_gw_res_ana}).  Note that the second line in Eq.~\eqref{eq:omega_gw_res_ana} is just a factor that takes into account the non-linear cutoff scale, and it reduces to 1 and 0 for $k \leq 2 k_\text{NL}/(1+\sqrt{3})$ and $k \geq 2 k_\text{NL}/\sqrt{3}$, respectively, smoothly interpolated in between.

\begin{figure}[ht] 
  \centering \includegraphics[width=1\columnwidth]{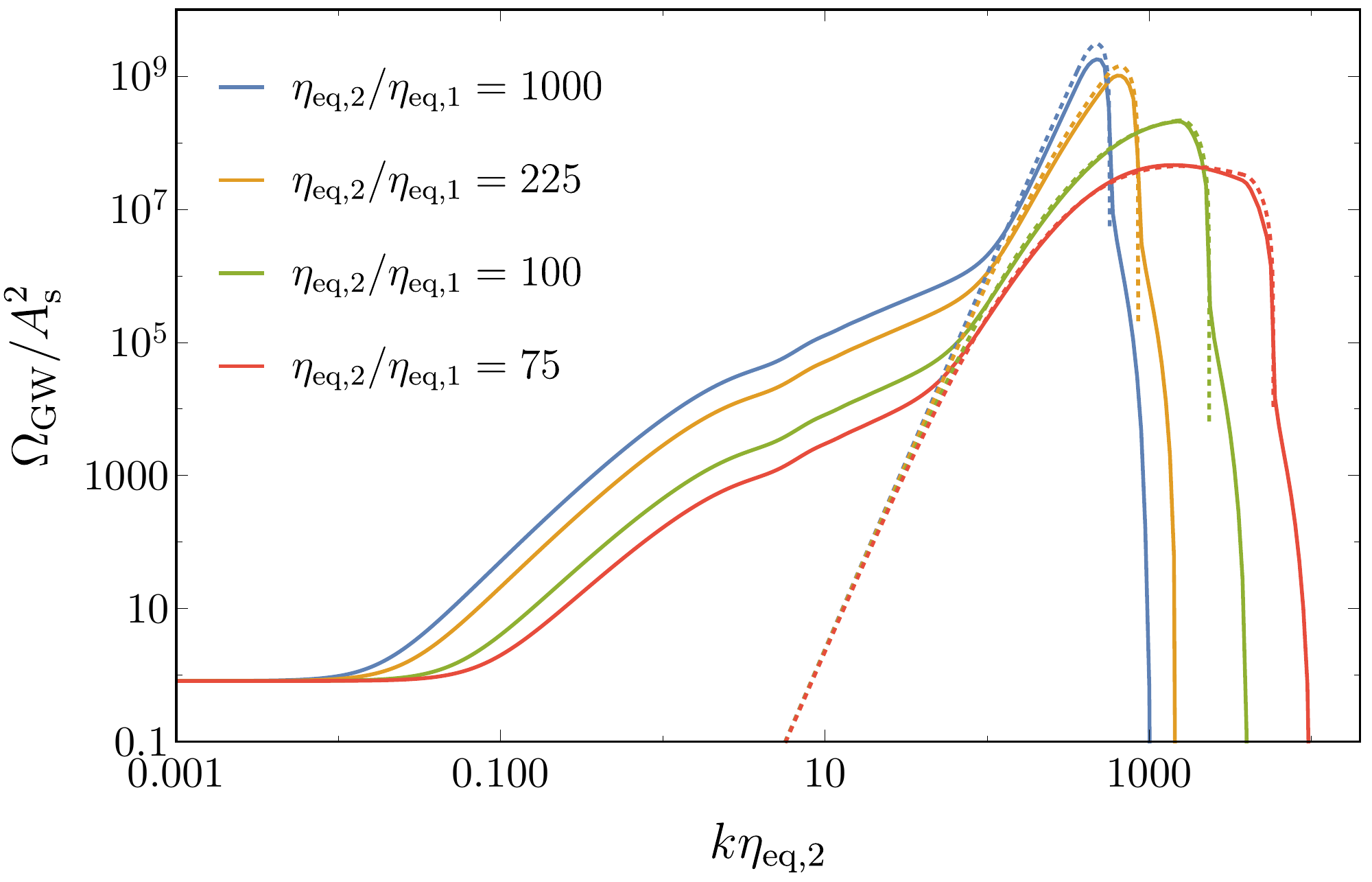} 
        \caption{
        The solid lines show the energy density parameters of the induced GWs normalized by $A_\text{s}^2$.
        We take $\eta_\eqs/\eta_\eqf = 1000\  (\text{top}), 225, 100$, and $75$ (bottom) and assume $n_\text{s} = 1$ for all plots.
        We also plot the approximate expressions for the resonance peak, given in Eq.~(\ref{eq:omega_gw_res_ana}), with dotted lines.
        }
        \label{fig:gw_spectrum_norm}
\end{figure}

In Fig.~\ref{fig:gw_spectrum_norm}, we show the numerical results of the energy density parameter of the induced GWs normalized by $A_\text{s}^2$. 
The spectrum has a unique shape.
For the case of the blue (top) line ($\eta_{\text{eq},2}/\eta_{\text{eq},1} = 1000$), the spectral shape is similar to the sudden transition limit in Ref.~\cite{Inomata:2019ivs}. 
For example, the spectral index has values  0, 3, 1, and 7, approximately in $k \eta_\text{eq,2} \lesssim 0.01$, $0.01 \lesssim k \eta_\text{eq,2} \lesssim 1$, $1 \lesssim k \eta_\text{eq,2} \lesssim 100$, and $100 \lesssim k \eta_\text{eq,2} \lesssim 500$, respectively (see Fig.~\ref{fig:gw_spectrum_norm}). 
The normalization in each range is, however, smaller than the sudden limit in Ref.~\cite{Inomata:2019ivs} since the evaporation takes finite time. 
Note that the peak scale is roughly the same as $k_\text{NL}$ for large $\eta_\eqs/\eta_\eqf$ ($=1000$, $225$, and $100$ in Fig.~\ref{fig:gw_spectrum_norm}). 
On the other hand, for small $\eta_\eqs/\eta_\eqf$ ($=75$ in the figure) in which the perturbation with $k_\text{NL}$ enters the horizon well before the PBH-dominated era, the peak scale is determined by the balance between the enhancement of the resonance and the suppression of $\Phi$ at the horizon entry during the eRD era because a smaller scale gravitational potential oscillates faster after the reheating but gets more suppressed at the horizon entry.

From Fig.~\ref{fig:gw_spectrum_norm}, we can also see that the shorter PBH-dominated era (smaller $\eta_\eqs/\eta_\eqf$) leads to the spectrum with a lower peak and a smaller-scale cutoff (large $k_\text{NL}$).
The peak height is lowered because the shorter PBH-dominated era makes the given mode entering the horizon deeper in the eRD era (larger $k \eta_\text{eq,1}$ for given $k \eta_\text{eq,2}$), which makes the normalization factor smaller (see Figs.~\ref{fig:s_low} and \ref{fig:knl_ns1}). The $k_\text{NL}$ becomes larger because the shorter PBH-dominated era delays the approach of the non-linearity. 
In summary, the deformation of the shape in the large $k$ side in Fig.~\ref{fig:gw_spectrum_norm} for smaller values of $\eta_\text{eq,2}/\eta_\text{eq,1}$ is due to the fact that the scalar source modes entered the horizon in the eRD era. 
We also plot Eq.~(\ref{eq:omega_gw_res_ana}) with dotted lines in Fig.~\ref{fig:gw_spectrum_norm}.
We can see that the analytical formula fits the numerical results well around the peak.

\section{Effects of finite-width mass function}
\label{sec:width}

So far, we have studied the GWs induced after the evaporation of \emph{monochromatic} PBHs.  It is, however, crucial to consider the effects of finite width of the PBH mass function since a broad mass spectrum of PBHs will not lead to a sufficiently sudden transition from the eMD era to the RD era.\footnote{Since the life time of a PBH depends on its spin, the finite width of the spin distribution of PBHs might affect the suddenness of the reheating, though the spin parameter is about a few percent for PBHs produced during a RD era~\cite{Mirbabayi:2019uph,DeLuca:2019buf}. We leave the discussion on the effect of the spin distribution for future work.}

Let us parametrize the initial mass function as follows:
\begin{align}
\rho_{\text{PBH,i}} =& \int \rho_\text{PBH,i}(M_\text{PBH,i}) \text{d} \ln M_\text{PBH,i} 
\nonumber \\
\simeq &  \int \rho_\text{PBH,i}(M_\text{PBH,i}(\eta_\text{eva})) \text{d} \ln \eta_\text{eva} .
\end{align}
In the last equality, we changed the variable for convenience for numerical calculations by using Eq.~\eqref{eq:mass-time_relation} and the relation $t_\text{eva} \propto \eta_\text{eva}^3$ valid in a MD era and by neglecting $\eta_\text{i} (\ll \eta_\text{eva})$.  

The energy density of the PBHs can be calculated by integrating the corresponding formula in the case of the monochromatic mass over the initial mass function, i.e.,
\begin{align}
\rho_\text{PBH} (\eta) 
\simeq & \int \rho_\text{PBH,i}(M_\text{PBH,i}(\eta_\text{eva})) \left( 1 - \left(\frac{ \eta }{\eta_\text{eva}} \right)^3 \right)^{\frac{1}{3}} \left( \frac{a(\eta_\text{i})}{a(\eta)} \right)^3  \nonumber \\
& \times  \Theta (\eta_\text{eva} - \eta )  \text{d} \ln \eta_\text{eva} .\label{rho_PBH_convolution} 
\end{align}
Here, we used $t \propto \eta^3$ and neglected $\eta_\text{i}$ against $\eta_\text{eva}$, so the expression is valid after the PBHs dominate the Universe.

It would be time-consuming to calculate this integral at each time with simultaneously solving the differential equation to determine $a(\eta)$. 
Therefore, we numerically record and interpolate the comoving PBH energy density $\rho_\text{PBH}(\eta) (a(\eta)/a(\eta_\text{i}))^3$ as a function of conformal time for a given parameter set.  Then, using it, we solve equations of motion for other quantities such as $\rho_\text{r}$ and perturbations.

 Note that the lifetime of a BH is $t_\text{eva} \sim M_\text{PBH,i}^3 \sim \eta_\text{eva}^3$, so $M_\text{PBH,i} \propto \eta_\text{eva}$.  This means, in particular, if we assume the log-normal distribution for the PBH masses~\cite{Kannike:2017bxn}, the distribution of $\eta_\text{eva}$ can also be written as the log-normal distribution with the same variance,
  \begin{align}
\rho_\text{PBH,i}(M_\text{PBH,i}(\eta_\text{eva})) = \frac{\rho_\text{PBH,i} }{\sqrt{2\pi} \sigma } \exp\left( - \frac{\left(\text{ln}(\eta_\text{eva}/\eta_\text{eva,0})\right)^2}{2 \sigma^2} \right) ,
\label{eq:pbh_mass_func_with_sigma}
  \end{align}
where $\eta_\text{eva,0}$ is the central value of the evaporation time, and $\sigma$ is its standard deviation.  
Note that the limit of the small variance ($\sigma \rightarrow 0$) corresponds to the monochromatic PBH mass function, which is discussed in the previous section. 
  Based on this log-normal distribution, we study the effect of a finite $\sigma$ numerically in the following.

\begin{figure}[tbh!]
\begin{center}
  \includegraphics[width = 1. \columnwidth]{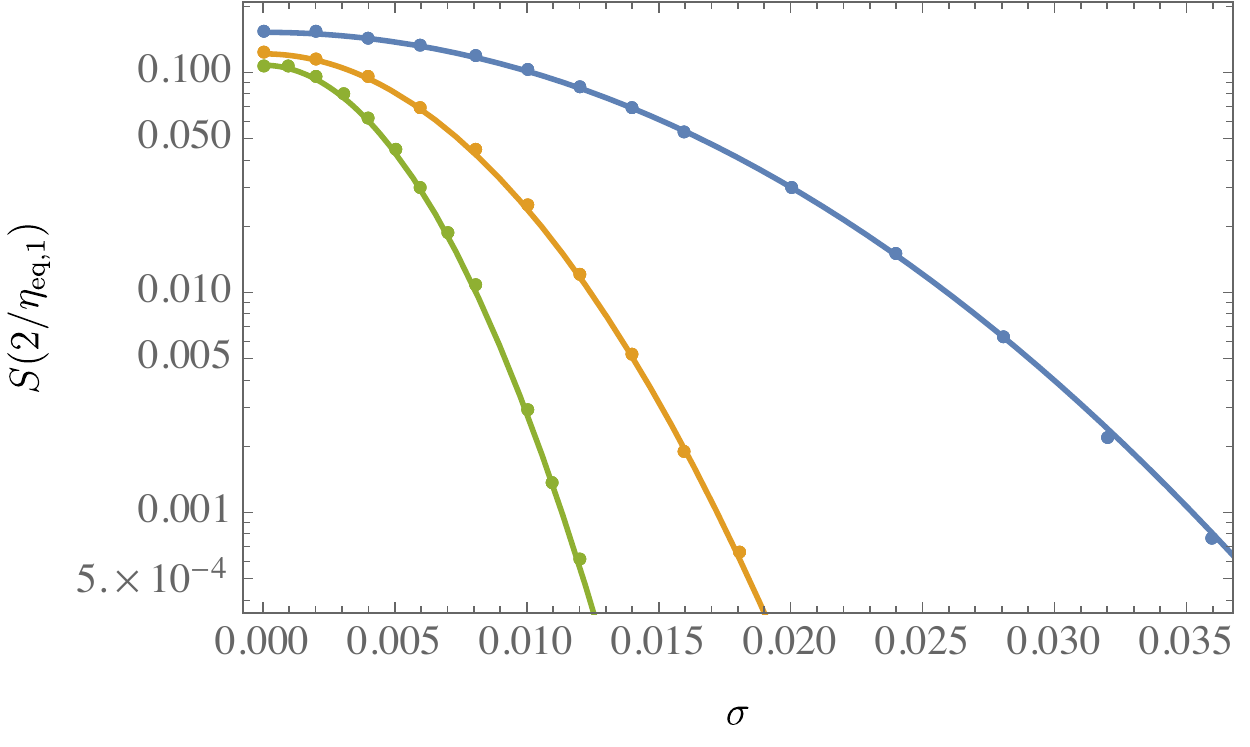}
      \caption{ \label{fig:sigma_dependence} Dependence of the normalization factor $S(k = 2/\eta_\text{eq,1})$ on the width of PBH mass function, which is parametrized by Eq.~(\ref{eq:pbh_mass_func_with_sigma}) with $\sigma$. The blue (top), orange (middle), and green (bottom) dots correspond to the case of $\eta_\text{eq,2}/\eta_\text{eq,1} = 75$, 150, and 225, respectively.  The lines are obtained by multiplying the values at $\sigma \rightarrow 0$ with a factor $\exp ( - (c \sigma k \eta_\text{eq,2})^2  )$. }
      \end{center}
\end{figure}

Similarly to Sec.~\ref{sec:evo_pertb}, we can study the normalization factor $S$, which is nothing but the value of the transfer function of the gravitational potential $\Phi$ just after the evaporation. Its dependence on $\sigma$ is shown as dots in Fig.~\ref{fig:sigma_dependence}.  These can be well fitted by
\begin{align}
S(k, \sigma) = & S(k) \exp \left( - \left(c \sigma k \eta_\text{eq,2}\right)^2 \right), \label{eq:sup_factor_with_sigma}
\end{align}
where $c^2 \simeq 0.18$, as shown by the solid lines in Fig.~\ref{fig:sigma_dependence}.  
We have also studied the $k$ dependence of the normalization factor with a fixed finite $\sigma$ and find the consistency with the above equation.
(Note that the exponential dependence on $\sigma$ may not be surprising since we introduced $\sigma$ as the standard deviation of $\text{ln} M_\text{PBH,i}$ rather than that of $M_\text{PBH,i}$.) 

This suppression in Eq.~\eqref{eq:sup_factor_with_sigma} can be understood as follows.
As discussed in Ref.~\cite{Inomata:2019zqy}, the GWs decouple from the source around when the time derivative and the wavenumber of the GW mode become comparable. This further implies that the induced GWs are suppressed if and only if the reheating transition time scale is longer than the time scale of the GW mode.\footnote{
TT thanks Misao Sasaki for illuminating discussion on this point.
}
 The former is $\sigma \eta_\text{eq,2}$ (for a small $\sigma (\ll 1)$) and the latter is $k^{-1}$ for the mode with its wavenumber $k$, so the criterion of non-suppression is $k \sigma \eta_{\text{eq,2}} \ll 1$.  This is nothing but what Eq.~\eqref{eq:sup_factor_with_sigma} tells us. 
 For a given $\sigma$, there should be an effective maximal $k$ that is not significantly suppressed by the effect of the finite width $\sigma$: $k_\sigma \equiv (c \sigma \eta_\text{eq,2})^{-1} $. The maximal $k$ that allows enhancement and the linear analysis is 
$
k_\text{max} \sim \min \left [  k_\text{NL},  k_\sigma  \right ].
$ 
Note that this implies that there will be no enhancement at all if $\sigma = \mathcal{O}(1)$ as $k_\text{max} \sim \eta_\text{eq,2}^{-1}$.

\begin{figure}[tbh!] 
  \centering \includegraphics[width=0.9\columnwidth]{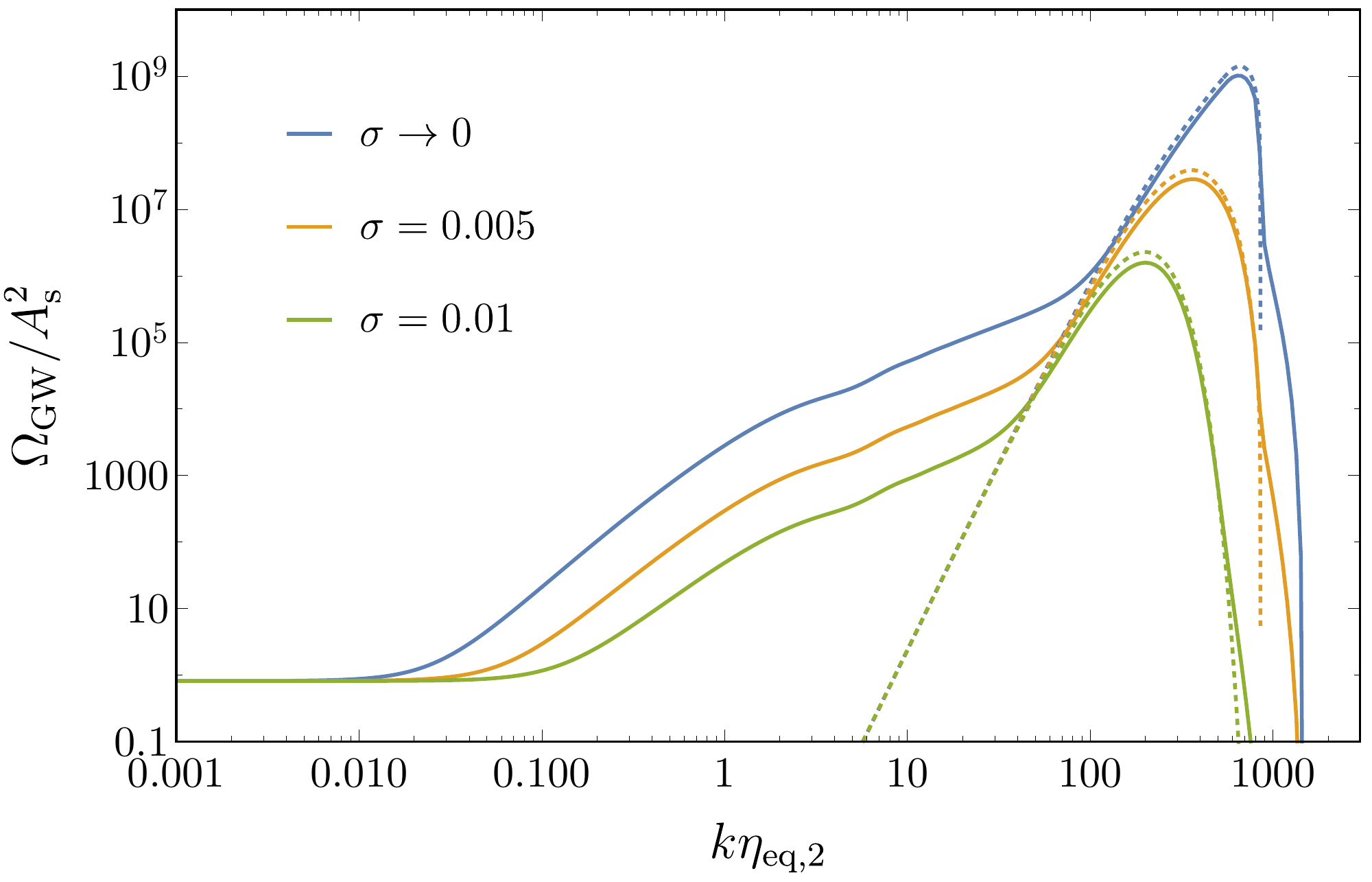}
        \caption{
        Dependence of the induced GWs on the width of PBH mass function, which is parametrized by Eq.~(\ref{eq:pbh_mass_func_with_sigma}) with $\sigma$.
        We take $\eta_{\text{eq},2}/\eta_{\text{eq},1} = 225$ for all lines and take $\sigma \rightarrow 0$ (top), $\sigma = 0.005$ (middle), and $0.01$ (bottom).
        We assume the same power spectrum of curvature perturbations as in Fig.~\ref{fig:gw_spectrum_norm}, which is given by Eq.~(\ref{eq:pzeta_almost_sc_inv}) with $n_\text{s} = 1$ (the plot for $\sigma \rightarrow 0$ is the same as the plot for $\eta_\eqs/\eta_\eqf = 225$ in Fig.~\ref{fig:gw_spectrum_norm}).
        We also show the analytic approximation formulas for the resonance peak, given in Eq.~(\ref{eq:omega_gw_res_w_sigma}), with dotted lines.
        }
        \label{fig:gw_profile_with_sigma}
\end{figure}

The solid lines in Fig.~\ref{fig:gw_profile_with_sigma} show the energy density parameters of induced GWs with different values of $\sigma$, which are numerically calculated with the approximation of $S(k, \sigma) \simeq S_\text{low}(k) \exp \left( - (c \sigma k \eta_\text{eq,2})^2 \right)$.
The dotted lines in Fig.~\ref{fig:gw_profile_with_sigma}  show the analytic approximation formula for the energy density parameter, which is given as
\begin{align}
	\Omega^{(\text{res})}_\text{GW}(\eta_\text{c},k, \sigma) = \Omega^{(\text{res})}_\text{GW}(\eta_\text{c},k, 0) G(z),
	\label{eq:omega_gw_res_w_sigma}
\end{align}
where $\Omega^{(\text{res})}_\text{GW}(\eta_\text{c},k, 0)$ is given by Eq.~(\ref{eq:omega_gw_res_ana}) and $G(z)$ is given with the parameter $z = (c \sigma k \eta_{\text{eq},2})^2$ as
\begin{align}
	G(z) = \frac{15 \ee^{-4z}}{64 z^{5/2}} \left(2\sqrt{z}\left(2z -3\right) + \left(4z^2 -4 z +3\right) \ee^z \sqrt{\pi} \text{Erf}\left(\sqrt{z}\right) \right),
\end{align}
where Erf denotes the error function. 
We explain the derivation of this approximation formula in Appendix~\ref{app:app_formula}.
From Fig.~\ref{fig:gw_profile_with_sigma}, we can see that the approximation formula fits the numerical results well.
Also, we see how small $\sigma$ significantly reduces the strength of the induced GWs. 

We do not show the results with $\sigma > 0.01$, such as $\sigma = 0.1$, in Fig.~\ref{fig:gw_profile_with_sigma}.
For example, in the case of $\sigma = 0.1$, we find that the maximal wavenumber for the enhancement ($k_\sigma$) becomes $k_\sigma \sim 20/\eta_\eqs$.
In this case, we cannot neglect the contribution during $\eta<\eta_\text{eva}$ in the function $I$, defined in Eq.~(\ref{eq:i_formula}), because the $k_\sigma$ is not far from $1/\eta_\text{eva}$.
Therefore, a more detailed analysis is needed to obtain the plots for $\sigma = \mathcal O(0.1)$, which is left for future work. 
Having said that, we have numerically confirmed that the energy density parameter for the induced GWs with $\sigma = 0.1$ is enhanced by one order of magnitude around the peak scale even if we neglect the contribution during $\eta<\eta_\text{eva}$ in the function $I$. 

\section{Constraints on PBH abundance} 
\label{sec:measure_pbh}

In previous sections, we have shown that the enhancement of the induced GWs can be caused by PBH evaporation.
Here, we discuss the constraints on the initial PBH abundance $\beta$ through measurements of the enhanced induced GWs. 

\begin{figure}[ht] 
  \centering \includegraphics[width=1\columnwidth]{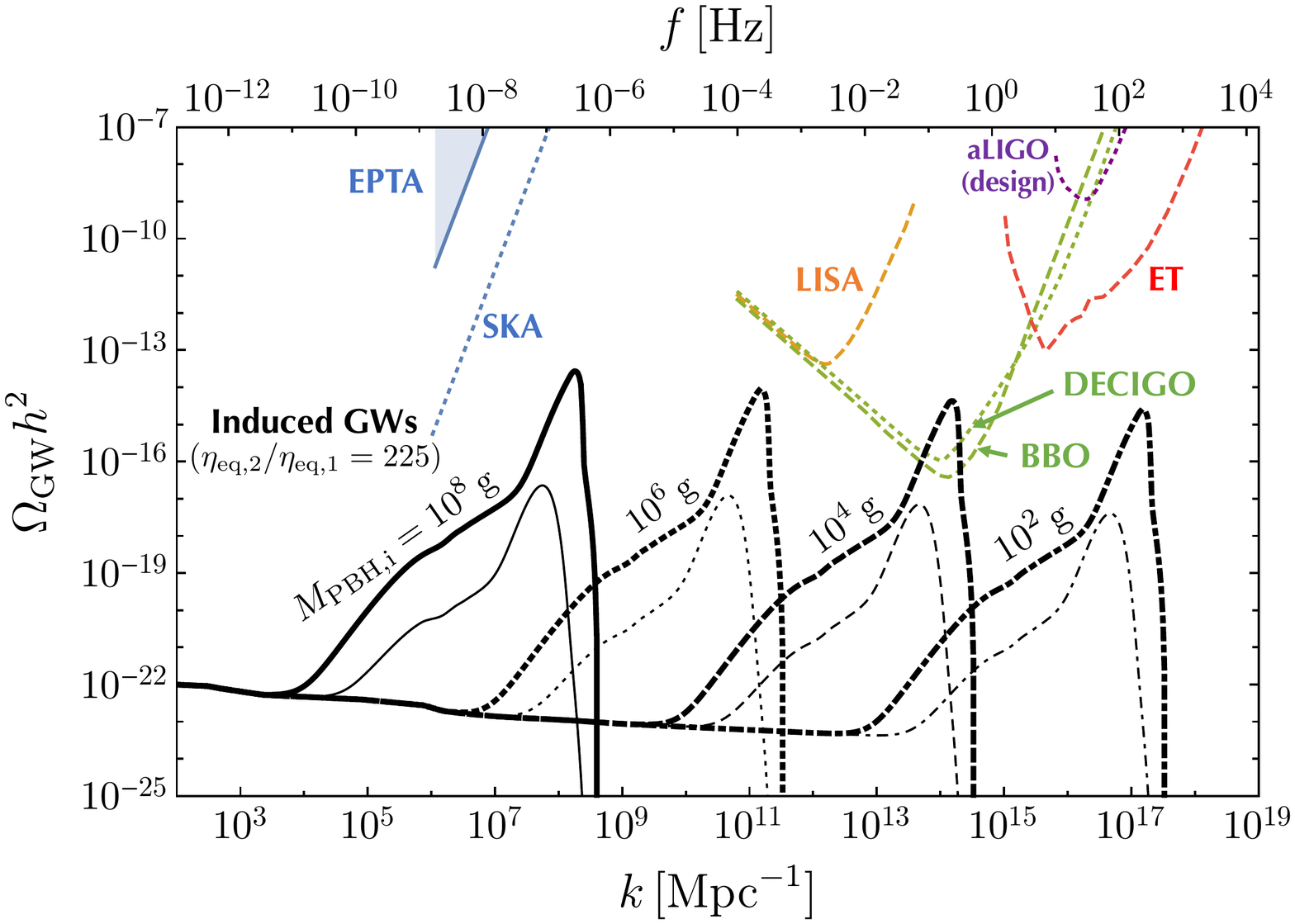} 
        \caption{
        The black lines show GW spectra with $\eta_{\text{eq},2}/\eta_{\text{eq},1} = 225$ and $M_{\text{PBH,i}} = 10^8$\,g (solid), $10^6$\,g (dotted), $10^4$\,g (dashed), and $10^2$\,g (dot-dashed).
        The thick black lines show results for the monochromatic mass function ($\sigma \rightarrow 0$) and the thin black lines show the results for $\sigma = 0.01$.
        The other lines show the future or current sensitivity curves (see Refs.~\cite{Inomata:2018epa,Inomata:2019ivs} for details).
        }
        \label{fig:gw_spectrum_erd}
\end{figure}

In Fig.~\ref{fig:gw_spectrum_erd}, we show the energy density parameters of the GWs induced by the scalar perturbations with $A_\text{s} = 2.1\times 10^{-9}$, $k_* = 0.05$\,Mpc$^{-1}$, and $n_\text{s} = 0.96$~\cite{Aghanim:2018eyx}.
Since the evaporation time $\eta_\text{eva}$ is determined by the initial mass of PBHs, as in Eqs.~(\ref{eq:tr_pbh_mass}) and (\ref{eq:kr_tr}), the peak scale of the induced GWs also depends on the PBH mass. 
From this figure, we can see that, even if the power spectrum of the curvature perturbations is almost scale invariant up to $k_\text{NL}$,\footnote{To produce a large enough number of tiny PBHs to realize the PBH-dominated era, large-amplitude scalar perturbations are required on the scale much smaller than the peak scale of the induced GWs.
In this sense, our assumption of the almost scale-invariant spectrum for $k\leq k_\text{NL}$ is conservative.} the induced GWs could be observed by future detectors, such as DECIGO, BBO, and LISA, depending on the mass function of PBHs.

As shown in Figs.~\ref{fig:gw_spectrum_norm}, \ref{fig:gw_profile_with_sigma}, and \ref{fig:gw_spectrum_erd}, the spectrum $\Omega_\text{GW}$ of the GWs induced after evaporation has a unique shape. 
On the largest scales, the slope is twice as that of the primordial curvature perturbations, $k^{2 (n_\text{s} -1)}$.  
This contribution is produced much after the evaporation. 
When $k$ becomes large, $\Omega_\text{GW}$ increases as $k^3$ up to $k_\text{eva} (\simeq k_\text{eq,2})$, and then becomes approximately $k^1$. 
Note that $k_\text{eva}$ has a simple dependence on the PBH mass, $k_\text{eva} \propto T_\text{R} \propto M_\text{PBH,i}^{-3/2}$ (see Eqs.~\eqref{eq:tr_pbh_mass} and \eqref{eq:kr_tr}).
The resonance peak starts with the steep slope $k^7$.
The larger $k$ side of the spectrum depends significantly on the period of the PBH-dominated era, which is controlled by $\beta$, and the width of the mass function. 
Note, however, that the spectral shape of the large $k$ side is not robust when we consider $k \simeq k_\text{NL}$.  Since $k_\text{NL}$ is the cutoff to avoid the nonlinearity of the density perturbations,  the value of $\Omega_\text{GW}$ should be regarded as a lower bound on the strength of the GWs rather than the precise prediction.

Following the same procedure as in Ref.~\cite{Inomata:2018epa}, we calculate the signal-to-noise ratio for each project and derive the region of $\beta$ that can be constrained by the future GW projects.
Here, we assume the power spectrum of curvature perturbations that is taken in Fig.~\ref{fig:gw_spectrum_erd} ($A_\text{s} = 2.1\times 10^{-9}$, $k_* = 0.05$\,Mpc$^{-1}$, and $n_\text{s} = 0.96$).
The signal-to-noise ratio is given by~\cite{Thrane:2013oya}
\begin{align}
SNR  = \sqrt{2T_\text{obs}} \left[ \int^{f_\text{max}}_{f_\text{min}} \dd f \, \left(\frac{\Omega_\text{GW}(f)}{\Omega_\text{GW,eff}(f)}\right)^2 \right]^{1/2},
\label{eq:rho_def}
\end{align}
where $T_\text{obs}$ is the observation time, and $(f_\text{min}, f_\text{max})$ is the range of observable frequencies for each project. 
$\Omega_\text{GW,eff}$ is the effective sensitivity curve for each project.
To show the potential of each observation, we assume the perfect subtraction of foreground here (see e.g.~Ref.~\cite{Cutler:2009qv} for subtraction techniques).\footnote{The extragalactic foreground from binary white dwarfs and main sequence stars might be difficult to be subtracted~\cite{Schneider:2000sg,Farmer:2003pa,Schneider:2010ks}. Since the foreground could contaminate the sensitivity curves in $f<\mathcal O(0.1)$Hz, the constraints on the abundance of PBHs with $M_{\text{PBH,i}} \gtrsim 10^4$\,g might be affected by the foreground. }
In Fig.~\ref{fig:gw_spectrum_erd}, we plot $\Omega_\text{GW,eff}h^2/\sqrt{T_\text{obs} f/10}$ as benchmark sensitivities of the future projects for stochastic GWs, where we take $T_\text{obs} = 18$\,years for EPTA, $T_\text{obs} = 20$\,years for SKA, and $T_\text{obs} = 1$\,year for the others as fiducial values (see Ref.~\cite{Inomata:2018epa} for details).\footnote{
Note that, even if there is no intersection between the GW spectrum and the sensitivity curves in Fig.~\ref{fig:gw_spectrum_erd}, the signal-to-noise ratio could be larger than unity because the ratio is defined as Eq.~(\ref{eq:rho_def}). This is why we regard the curves as \emph{benchmark} ones.}
To save the computational time, we use the analytical formulas given in Eqs.~(\ref{eq:omega_gw_res_ana}) and (\ref{eq:omega_gw_res_w_sigma}).
Since the peak scale and height of the GW spectrum are determined by the mass of PBHs and the length of the PBH-dominated era, which is parametrized by $\eta_\eqs/\eta_\eqf$,
 we first obtain the minimum value of $\eta_\eqs/\eta_\eqf$ which makes the signal-to-noise ratio unity ($SNR = 1$) for each PBH mass and each observation.\footnote{Strictly speaking, the increase of $\eta_\eqs/\eta_\eqf$ does not necessarily mean the increase of $SNR$ at least in our analysis. This is because the smaller $\eta_\eqs/\eta_\eqf$ leads to the smaller cutoff scales (see Fig.~\ref{fig:gw_spectrum_norm}). 
 However, in this paper, we assume for simplicity that the induced GWs from the non-linear perturbations ($k>k_\text{NL}$) should be larger than the induced GWs from the linear perturbations in the case of the shorter PBH-dominated era, which leads to the smaller cutoff scales. This is why we consider the \emph{minimum} value of $\eta_\eqs/\eta_\eqf$ to realize $SNR=1$.}
After that, using Eq.~(\ref{eq:beta_eta_rel}), we derive the curves for the PBH abundance ($\beta$) that can be probed by future observations. 

\begin{figure}[tbh!] 
  \centering \includegraphics[width=0.9\columnwidth]{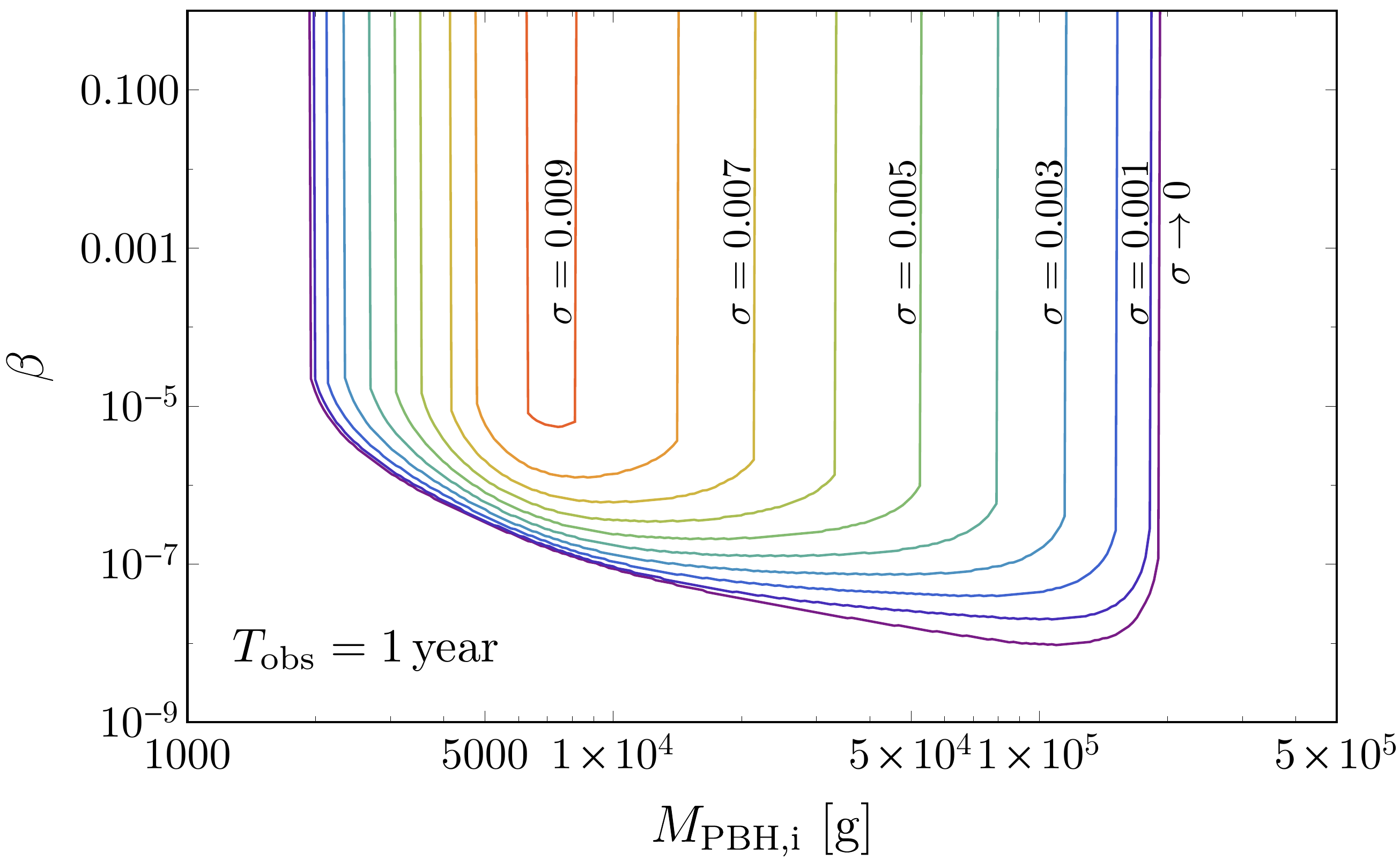}
        \caption{
        The parameter regions that can be investigated by one-year observation of DECIGO with different values of the width $\sigma$. 
        In the regions above the lines, the GWs can be measured.
        We assume the almost scale-invariant power spectrum of curvature perturbations (see the text).
        The outermost line shows the result for $\sigma \rightarrow 0$ and the difference of $\sigma$ between two adjacent lines is $0.001$.
        We omit lines for BBO to make this figure simple, but the results are almost the same.
 		We take the signal-to-noise ratio as unity for all lines ($SNR=1$).
 		Note that LISA cannot investigate the abundance of the tiny PBHs with its one-year observation in this setup.
        }
        \label{fig:beta const_sigma_1yr}
\end{figure}

\begin{figure}[tbh!] 
  \includegraphics[width=0.9\columnwidth]{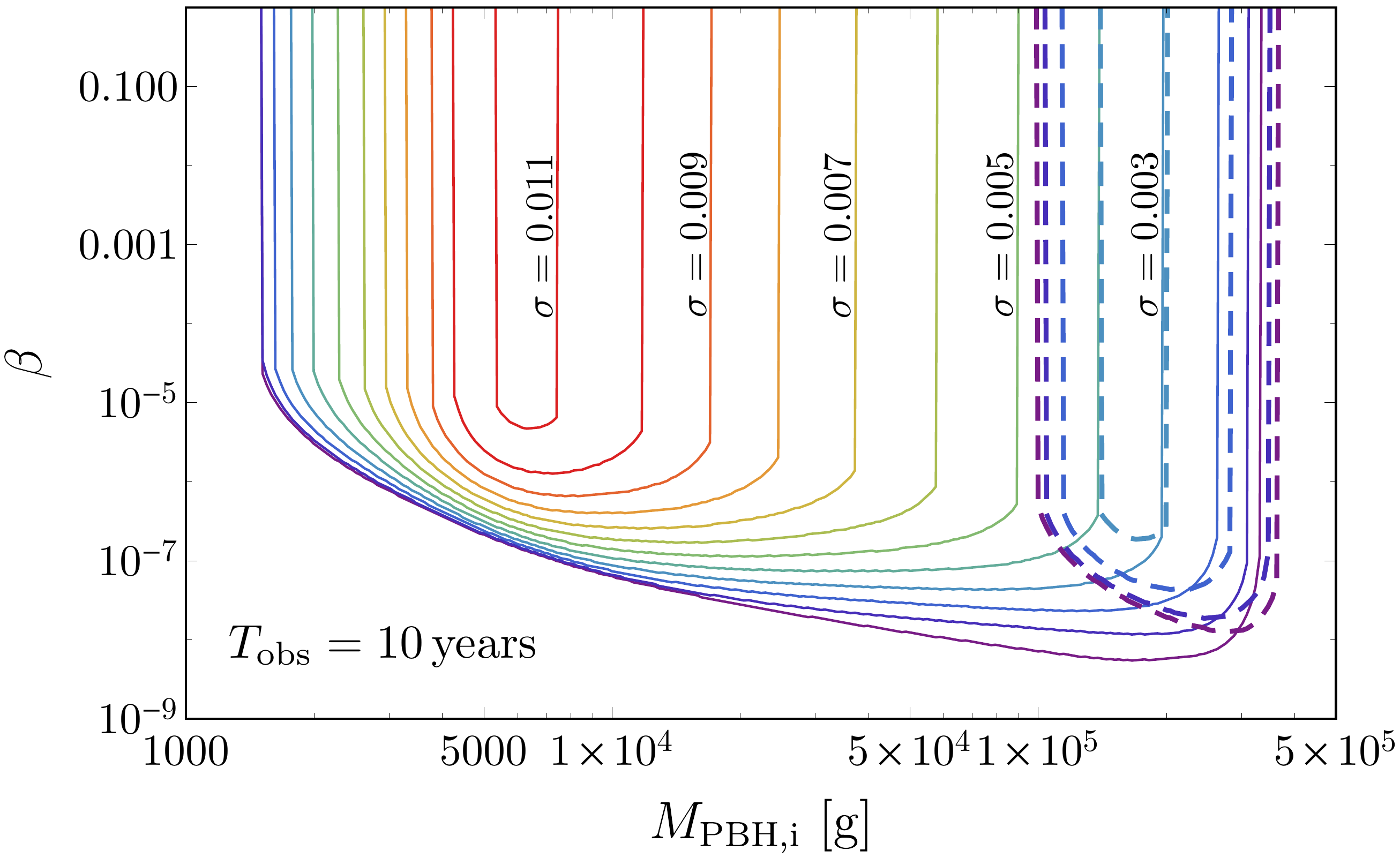}
        \caption{
        The parameter regions that can be investigated by 10-year observation of DECIGO and LISA.
        The solid lines show the regions for DECIGO and the dashed lines show the regions for LISA.
        Note that the same color indicates the same value of $\sigma$ regardless of the types of the lines.
        Except for the observation time, we take the same parameters as in Fig.~\ref{fig:beta const_sigma_1yr}.
        }
        \label{fig:beta const_sigma_10yrs}
\end{figure}

Figures~\ref{fig:beta const_sigma_1yr} and~\ref{fig:beta const_sigma_10yrs} show the results of the signal-to-noise-ratio analysis for $T_\text{obs} = 1$\,year and 10\,years, respectively.
From these figures, we can see that if the mass function is narrow as $\sigma \lesssim 0.01$, the future observations could constrain the initial PBH fraction, $\beta$, even for the almost scale-invariant spectrum of curvature perturbations.
In the narrow limit of the mass function ($\sigma \rightarrow 0$), the future projects could constrain the abundance of the PBHs with $\beta>\mathcal O(10^{-5}-10^{-8})$ in $2\times 10^3$\,g $\lesssim M_\text{PBH,i} \lesssim 4 \times 10^5$\,g. 
For example, the one-year observation of DECIGO can put the upper bound on $\beta$ as $\beta \lesssim 10^{-7} (M_\text{PBH,i}/10^4 \text{g})^{-1}$ in $2\times 10^3$\,g $\lesssim M_\text{PBH,i} \lesssim 2 \times 10^5$\,g.
The PBH mass dependence of the upper bound on $\beta$ can be understood as follows.
The amplitude of the induced GWs is mainly determined by the parameter $\eta_\eqs/\eta_\eqf$, which is only related to $\beta/\beta_\text{min}$ (see Eq.~(\ref{eq:beta_eta_rel})) and the parameter $\beta_\text{min}$ is proportional to $M_{\text{PBH},\text{i}}^{-1}$ (see Eq.~(\ref{eq:beta_min_def})).
Then, the upper bound has the mass dependence $\beta \propto M_\text{PBH,i}^{-1}$.

Note that if we consider the blue-tilted power spectrum of the curvature perturbations, the future observations could detect the induced GWs that are related to the PBH evaporation with a wider mass function. This is because the enhancement should exist as long as $\sigma < 1$ holds and because the amount of the induced GWs also depends on the amplitudes of the curvature perturbations on small scales.

\section{Discussions and Conclusions}
\label{sec:conclusions}

In this paper, we have shown that scalar perturbations can produce a large amount of GWs soon after the evaporation of the tiny PBHs ($M_\PBH <10^9$\,g) if the PBH mass function is sufficiently narrow.
The enhancement of the induced GWs occurs if the PBHs come to dominate the Universe by the time of its evaporation.
The Universe is dominated by PBHs in the early Universe if the initial fraction of PBHs in the total energy density is large enough at their production (see Eq.~\eqref{eq:beta_min_def}), and the PBH-dominated era ends with the Hawking evaporation of the PBHs.
We have carefully taken into account the evolution of the gravitational potential, source of the induced GWs, during the transition from the PBH-dominated era to the RD era and calculated the amplitude of the induced GWs.
As a result, we have found that the induced GWs can be observed by future detectors, such as DECIGO, BBO, and LISA, depending on the PBH mass function.

A physical picture of this phenomenon, the Poltergeist mechanism, is given as follows. 
In the PBH-dominated era, the fluctuation of the number density of PBHs grows since the pressure in the Universe is negligible.  These fluctuations are converted to the sound waves on the thermal bath by the reheating due to PBH evaporation.  Then, the oscillations of the sound waves and the associated oscillations of the gravitational potential produce the GWs by the resonance effect as the dominant production channel.  
Here, it is crucial to have an unsuppressed amplitude for the gravitational potential (enhanced amplitudes for the sound waves) to produce the enhanced GWs, and this is realized by the sudden reheating transition. 
The evaporation rate of the PBH increases as time goes by because the Hawking temperature of the PBH is proportional to the inverse of its mass, which leads to a rapid instability and realizes the sudden reheating transition. 
The sudden transition prevents the otherwise large suppression of the gravitational potential during the transition and leads to the fast oscillations of the gravitational potential with the amplitude not suppressed much. 
The fast oscillations of the gravitational potential enhance the induced GWs. 
Note that the degree of suddenness strongly depends on the width of PBH mass function, and the narrow mass function is required for the induced GWs to be detectable by the future observations.

Another interesting point of the GWs produced by the Poltergeist mechanism is that the typical frequency is not directly determined by the size of a single PBH, but it is determined by the typical wavelength of the fluctuations which enters the horizon by the end of the reheating implying a macroscopic scale of a fluid composed of many PBHs.
This scale is in turn controlled by the reheating temperature ($T_\text{R} \propto M_\text{PBH,i}^{-3/2}$) and the initial abundance of PBHs ($\beta$).  Thus, the frequency is small enough to be probed by the future GW observations for the appropriate range of the PBH mass, $\mathcal{O}(10^{3} \text{--} 10^5 )$ g.

We have also discussed the possibility that the future GW observations can probe the abundance of the tiny PBHs.
As a concrete example, we have discussed the case where the power spectrum of the curvature perturbations is almost scale invariant even at small scales.
Then, we have found that the initial PBH fraction can be measured or constrained if the PBH mass function is narrow ($\sigma \lesssim 0.01$).
We have also found that, in the narrow limit ($\sigma \rightarrow 0$), the initial fraction with $\beta>\mathcal O(10^{-5}-10^{-8})$ in $2\times 10^3$\,g $\lesssim M_\text{PBH,i} \lesssim 4 \times 10^5$\,g can be probed by the future observations. 
In particular, the one-year observation of DECIGO can constrain $\beta$ as $\beta \lesssim 10^{-7}(M_\text{PBH,i}/10^4\,\text{g})^{-1}$ in $2\times 10^3$\,g $\lesssim M_\text{PBH,i} \lesssim 2 \times 10^5$\,g.
See Figs.~\ref{fig:beta const_sigma_1yr} and \ref{fig:beta const_sigma_10yrs} for the cases with the finite width of the mass function.
Note again that the detectability also depends on the power spectrum of the curvature perturbations on small scales.  For example,   we can probe the evaporating PBHs with wider mass functions (within the range $\sigma \lesssim 1$) if the power spectrum of the curvature perturbations is blue-tilted.  

Here, let us discuss possibilities to realize a narrow width of the mass function.
A refined criterion for the PBH formation has recently been proposed in Refs.~\cite{Suyama:2019npc, Germani:2019zez}.  
It predicts a narrow mass function $f(M)$ even if we take into account the coarse graining by a window function and the effects of critical collapse~\cite{Germani:2019zez} provided that the power spectrum $\mathcal{P}_\zeta (k)$ of curvature perturbations is narrow. 
A possibility to produce a narrow $\mathcal{P}_\zeta(k)$ for $M_{\text{PBH,i}} \sim 10^4$~g is to utilize a parametric resonance. See e.g.~Refs.~\cite{Kawasaki:2006zv,Kawaguchi:2007fz, Cai:2018tuh, Martin:2019nuw, Cai:2019bmk, Chen:2020uhe} towards such a direction. 

We point out that there is a limitation on the narrowness of the width of $\mathcal{P}_\zeta (k)$.
To be concrete, we parametrize $\mathcal P_\zeta$ around the peak scale related to PBH production as $\mathcal P_\zeta = A_\zeta /(\sqrt{2\pi} \sigma_\zeta) \exp(-(\text{ln}(k/k_\text{peak}))^2/(2 \sigma_\zeta^2)$.
In a RD era, $A_\zeta \sim \mathcal O(10^{-2})$ is needed for the production of a sizable amount of PBHs when $\sigma_{\zeta}$ is $\mathcal{O}(1)$. 
 More precisely, the amount of PBHs depends on an \emph{integral} of $\mathcal{P}_\zeta (k)$ over $\text{ln} k$ with some window function (see e.g.~Ref.~\cite{Ando:2018qdb}).  
 To make the width narrower while keeping the fixed abundance of PBHs, the height of the peak of $\mathcal{P}_\zeta$ ($A_\zeta$) must become larger.  
 This implies that its value exceeds unity if we consider $\sigma_\zeta \lesssim 10^{-2}$, which leads to a non-perturbative and highly quantum regime of inflaton/curvature perturbations.  
 In this sense, there is a lower bound on $\sigma_\zeta$ for the standard analysis to be valid.\footnote{This discussion does not directly apply to the case of PBH formation in an eMD era or the era with a soft equation of state.  In the absence of the radiation pressure, the PBH formation threshold is lowered, so the above fundamental limit on the width becomes weaker than in the case of PBH formation in a RD era.}
It is desirable to find a relation between such a restriction on $\sigma_\zeta$ and that on $\sigma$, which is the width of the PBH mass function,  using the refined PBH formation criterion.

In addition to the initial mass function, we have to consider its potential time evolution via mergers of PBHs. We have studied these effects in Appendix~\ref{app:merger} considering the mergers of binary PBHs formed in the eRD era.  With the approximation of constant PBH mass during the merger process and with the assumption that the merger rate formula in Refs.~\cite{Sasaki:2016jop, Sasaki:2018dmp} is valid for the PBH-dominating ($f_\text{PBH} =1$) case, which is questioned in Ref.~\cite{Raidal:2018bbj}, we have found that the effects of mergers may not be negligible if $\beta \gtrsim 10^{-5}$ for the PBH masses of our interests.  We have not included such effects in Figs.~\ref{fig:beta const_sigma_1yr} and \ref{fig:beta const_sigma_10yrs} since our analyses on the merger processes are incomplete. That is, the constant-mass approximation may not be justified in the PBH-dominating scenario, and the merger rate formula may not be valid due to disruption of the binary by surrounding PBHs.  Nevertheless, the analyses imply that the constraints on the parameter space with large $\beta$ may not be valid.  The extension to the time-dependent mass case and the $N$-body simulation of the tiny PBHs should be done elsewhere.

Even if we take into account all these caveats on the width of the PBH mass function and the (potential) effects of mergers of PBHs, the amount of GWs we have discussed in this paper can still be larger in the observationally relevant frequency range than those from known GW production mechanisms in the PBH-dominating scenario (cf.~Fig.~\ref{fig:gw_summary_appendix} in Appendix~\ref{app:various_GWs}).
The future GW observations can constrain a part of the parameter space of tiny PBHs that evaporated in the early Universe ($2\times 10^3\, \text{g} \lesssim M_\text{PBH,i} \lesssim 4\times 10^5\,\text{g}$, $\beta>\mathcal O(10^{-5}-10^{-8})$), which has never been probed by any other robust means.

Finally, let us discuss several implications of this scenario for cosmology and particle physics.
Possibilities for dark matter are restricted.  At the final stage of the evaporation, any particle species in the theory are produced.  For PBH masses of $\mathcal{O}(10^3 \text{--} 10^5) \, \text{g}$, which can be probed by future observations, the abundance of any massive stable particles below the Planck scale exceeds the dark matter abundance~\cite{Hooper:2019gtx}.
Without an additional dilution  factor, massive dark matter candidates are excluded.  In this context, an axion(-like particle) is a good candidate for dark matter.  They are produced as relativistic particles by the Hawking radiation, which can be consistent with the effective neutrino species bound~\cite{Hooper:2019gtx}, while the dark matter abundance can be explained by the misalignment mechanism. 
Note also that the entropy production due to the evaporation can dilute unwanted extended objects such as magnetic monopoles predicted in the grand unified theories, which are expected not to be produced substantially via the evaporation (see a related comment in footnote~\ref{fn:pbh_catalyst} and Ref.~\cite{Fujita:2014hha}).

Interestingly, the tiny PBHs around the above parameter space, $M_\text{PBH,i} \sim \mathcal O(10^3 \text{--} 10^5)\,\text{g}$ and $\beta>\mathcal O(10^{-5} \text{--} 10^{-8})$, allow the generation of baryon asymmetry via leptogenesis in several ways: 
(i) thermal leptogenesis~\cite{Fukugita:1986hr} with right-handed Majorana neutrino somewhat heavier than the standard, $M_{\text{R}} \sim T_{\text{i}} (\sim \mathcal O(10^{13}  \text{ -- } 10^{14})\,\text{GeV})$ (see e.g.~Eq.~\eqref{eq:mpbh_k}), supplemented by the dilution of the baryon asymmetry due to the entropy production through the PBH evaporation, where the dilution factor is less than $\mathcal O(10^{-2})$ (see Eq.~(\ref{approximate_dilution})), 
(ii) TeV-scale thermal resonant leptogenesis after the evaporation with $T_\text{R} \sim \mathcal O(10^{3}  \text{ -- } 10^{6})\,\text{GeV}$ (see Eq.~\eqref{eq:tr_pbh_mass})~\cite{Pilaftsis:1997jf,Pilaftsis:2003gt}, 
and (iii) non-thermal (resonant) leptogenesis via the PBH evaporation with $T_\text{PBH} \sim \mathcal O(10^{8}  \text{ -- } 10^{10})\,\text{GeV}$ (see Eq.~\eqref{eq:t_pbh_m_pbh})~\cite{Baumann:2007yr,Fujita:2014hha}.
In addition, the evaporating PBHs could relax the Hubble tension~\cite{Hooper:2019gtx,Nesseris:2019fwr,Lunardini:2019zob} and also they are predicted in some inflationary and particle physics models~\cite{GarciaBellido:1996qt,Clesse:2015wea,Kawasaki:2015ppx,Linde:2012bt,Domcke:2020zez,Martin:2019nuw,Martin:2020fgl}.
From this perspective, our result could be a key to elucidate the mysteries of modern Cosmology in the near future.

\acknowledgments 
\noindent
We thank Hiromasa Nakatsuka for useful comments on the width of PBH mass function.
KI and TT thank Kazunori Kohri and Tomohiro Nakama for collaborations in the previous work.
TTY thanks Kavli IPMU for their hospitality during the corona virus outbreak.
This work was supported in part by World Premier
International Research Center Initiative (WPI Initiative), MEXT,
Japan, the JSPS Research Fellowship for Young Scientists (KI and TT),
JSPS KAKENHI Grants No.~JP18J12728 (KI), No.~17H01131 (MK), No.~17K05434
(MK),  No.~JP17J00731 (TT), No.~16H02176 (TTY), No.~17H02878 (TTY), and No.~19H05810 (TTY), 
MEXT KAKENHI Grant No.~15H05889 (MK),
the China Grant for Talent Scientific Start-Up Project (TTY), 
Advanced Leading Graduate Course for Photon Science (KI), 
IBS under the project code, IBS-R018-D1 (TT), and 
the Deutsche Forschungsgemeinschaft under Germany's Excellence Strategy - EXC 2121 Quantum Universe - 390833306 (KM).

\appendix


\section{Various GW sources in the PBH-dominating scenario}
\label{app:various_GWs}

\begin{figure}[tbh!] 
        \centering \includegraphics[width=1\columnwidth]{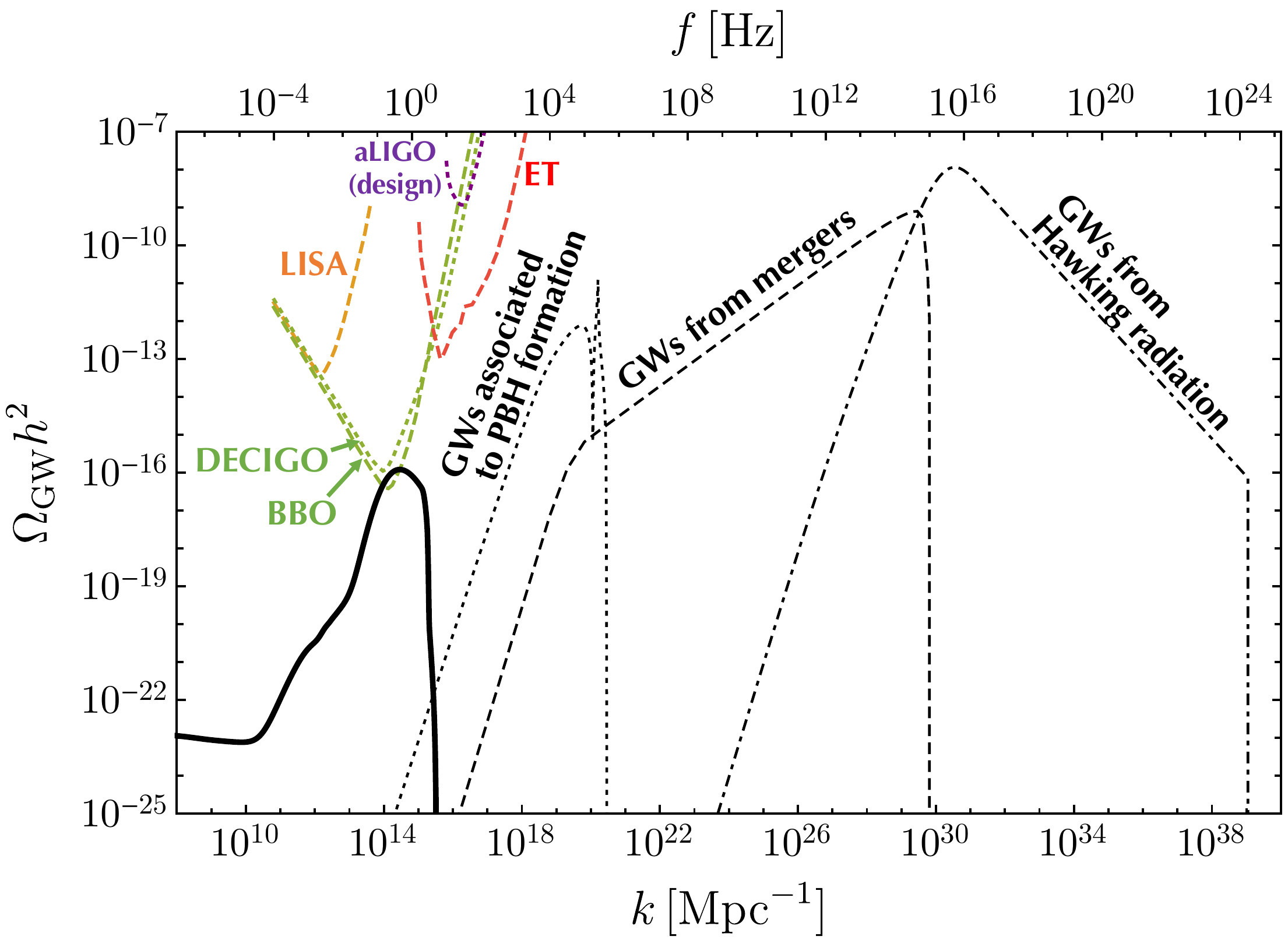}
        \caption{
        Summary of the GWs produced in the PBH-dominating scenario.
        $M_{\text{PBH,i}} = 10^4\,\text{g}$ and $\beta = 10^{-7}$ are taken for all plots.
        The thick black line shows the spectrum of the GWs induced just after the evaporation with $\sigma \rightarrow 0$, which we have focused on throughout this paper.
        The thin black lines show the GWs from the other sources, introduced in Appendix~\ref{app:various_GWs}.
       	Note that we assume $\mathcal P_\zeta = A_\zeta /(\sqrt{2\pi} \sigma_\zeta) \exp(-(\text{ln}(k/k_i))^2/(2 \sigma_\zeta^2)$ with $A_\zeta = 0.01$ and $\sigma_\zeta = 0.01$ for the spectrum of the GWs associated to PBH formation (black dotted).
        }
        \label{fig:gw_summary_appendix}
\end{figure}

The main topic of this paper is the second-order (scalar-induced) GWs produced just after the PBH evaporation. 
 In the present scenario, where PBHs dominate the Universe and evaporate, there are other sources of GWs.  
In this appendix, we introduce three types of GWs other than the GWs induced right after the PBH evaporation.
 The first one is the second-order (scalar-induced) GWs associated to the PBH {\em formation} scale, rather than the evaporation scale.  The second one is the gravitons emitted by the Hawking radiation process.  The third one is the GWs from many binary PBH merger events, which is separately discussed in Appendix~\ref{app:merger} in detail.  
 We summarize the spectrum of the three types of GWs in Fig.~\ref{fig:gw_summary_appendix}.
 Other various sources of GWs are comprehensively discussed in Ref.~\cite{Dolgov:2011cq}.  
 Note that these various GW components have different energy spectra, so the scenario is in principle highly predictive although it will be hopeless to observe these high-frequency GWs with technologies available in the near future.

\subsection{GWs associated to PBH formation}

Let us discuss the induced GWs associated to the PBH formation scale~\cite{Saito:2008jc,Saito:2009jt} first. To produce a substantial amount of PBHs, the curvature perturbations are assumed to be enhanced on the corresponding scale.  On this scale, the GWs are induced by the enhanced scalar perturbations.  
Taking into account the presence of the PBH-dominated era, the corresponding wavenumber is given by Eq.~\eqref{k_i}.

The energy density parameter of the induced GWs associated with the PBH formation events decreases during the PBH-dominated era, so the current magnitude of the GWs is given by Eq.~\eqref{eq:current_Omega_GW} times the following dilution factor:
\begin{align}
D\equiv  \frac{1}{a_\text{eva}/a_\eqf + 1} \simeq &  \left( \frac{g_{s*}(T_\text{R})}{g_{s*}(T_\text{eq,1})  } \right)^{\frac{1}{3}} \left( \frac{\beta}{\beta_\text{min}} \right)^{-\frac{4}{3}}, \label{approximate_dilution}
\end{align}
where we have assumed $a_{\text{eq},1} \ll a_\text{eva}$ in the second (almost) equality.
Except for this dilution factor, the formalism of the calculation is essentially the same as that used in Sec.~\ref{sec:induced_gws}. 
 The resultant GW spectrum is shown as the black dotted line in Fig.~\ref{fig:gw_summary_appendix} assuming the primordial curvature perturbations  $\mathcal P_\zeta = A_\zeta /(\sqrt{2\pi} \sigma_\zeta) \exp(-(\text{ln}(k/k_i))^2/(2 \sigma_\zeta^2)$ with $A_\zeta = 0.01$ and $\sigma_\zeta = 0.01$. 
 
 Let us discuss the spectral features of the GWs associated to PBH formation.
 Since we assume the widths of the mass function and the curvature perturbations are small, the spectral shape is similar to the case of the delta function source $\mathcal{P}_\zeta (k) \sim \delta (\text{ln}(k/k_\text{i}))$~\cite{Ananda:2006af}. 
 However, the infrared scaling $\Omega_\text{GW} \propto k^3$ is qualitatively different from that in the delta function case $\propto k^2$ because of the small but finite width~\cite{Espinosa:2018eve, Cai:2019cdl}.

 The dependence of the peak position and height of the spectrum on the mass $M_\text{PBH,i}$ and the abundance $\beta$ of PBHs is as follows.  The characteristic scale $k_\text{i}$ is proportional to $M_\text{PBH,i}^{-5/6} \beta^{-1/3}$ as shown in Eq.~\eqref{k_i}.  This is different from the standard relation $k_\text{i} \propto M_\text{PBH,i}^{-1/2}$ in the absence of the PBH-dominated era.  
 If we fix $A_\zeta$, the height of the induced GWs is insensitive to the mass and the abundance of PBHs except for the dilution factor in Eq.~\eqref{approximate_dilution}.
 Because of the dilution, the height of the peak scales as $M_\text{PBH,i}^{-4/3} \beta^{-4/3}$.
Note that, strictly speaking, $A_\zeta$ slightly depends on $\beta$, which leads to a small modification of the $\beta$ dependence of the peak height.

\subsection{GWs from Hawking radiation}
Gravitons emitted by Hawking radiation constitute stochastic GW background.  This contribution has been studied in the literature~\cite{Anantua:2008am, Dolgov:2011cq, Dong:2015yjs}. 
The GW spectrum is obtained by integrating the emitted graviton energy spectrum~\cite{Anantua:2008am}, which is the Planck distribution up to a gray-body factor:
\begin{align}
\frac{\text{d}^2 E}{\text{d}t\text{d}k} = \frac{\mathcal G \,  g_{\text{H}*, g} M_\text{PBH}^2}{32 \pi^3 M_\text{Pl}^4}\frac{k^3}{\ee^{k/T_\text{PBH}} - 1},  \label{Planck-dist}
\end{align}
where $g_{\text{H}*,g} \simeq 0.1$~\cite{Hooper:2019gtx} is the effective number of degrees of freedom for gravitons. 

In this subsection, we assume a monochromatic mass for PBHs since a small width such as $\sigma = 0.01$ would not lead to a qualitatively different result for GWs from Hawking radiation in contrast to the cases of the GWs associated to the PBH formation and the GWs produced right after the PBH evaporation. 

Just after the evaporation, $\Omega_\text{GW}(t,k)$ is calculated as
\begin{align}
&\Omega_\text{GW}(t_\text{eva}, k) 
\nonumber \\
&= \frac{k  n_\text{PBH}(t_i)}{a(t_\text{eva}) \rho_\text{tot}(t_\text{eva})} \left(\frac{a(t_\text{i})}{a(t_\text{eva})} \right)^3  \int_{t_\text{i}}^{t_\text{eva}}  \frac{\text{d}^2 E_\text{e}}{\text{d}t_\text{e} \text{d} k_\text{e}}(t_\text{e},k_\text{e})  \text{d} t_\text{e}, \label{Omega_GW_evaporation}
\end{align}
where the subscript e denotes the emission time of the gravitons, and $k_\text{e} = k / a(t_\text{e})$ is the physical momentum at emission.
The strength of the current GWs is obtained by using Eq.~\eqref{eq:current_Omega_GW} with identifying $\Omega_\text{GW}(t_\text{eva})$ with $\Omega_\text{GW}(t_\text{c})$.

The above integral is approximately performed as follows.  It is convenient to define a dimensionless integration variable $\kappa = k/ (a(t)T_\text{PBH}(t))$. This behaves as $\kappa \sim t^{-1/2}$, $t^{-2/3}$, and $(1-(t/t_\text{eva}))^{1/3}$ for $t \ll t_\text{eq,1}$ (during the eRD era), $t_\text{eq,1} \ll t \ll t_\text{eva}$ (during the eMD era), and $t \lesssim t_\text{eva}$ (close to the reheating), respectively.  Noticing that the integrand is dominated around $\kappa \simeq 1$, we may approximate it using the above approximation for $k/a_\text{eq,1} \ll T_\text{PBH,i}$, $k/a_\text{eva} \ll T_\text{PBH,i} \ll k/a_\text{eq,1}$, and $T_\text{PBH,i} \ll k/a_\text{eva}$, respectively. Up to the overall factor, $\mathcal G \, g_{H*,g}(T_\text{PBH}) T_\text{PBH,i} n_\text{PBH}(t_\text{eva}) / (32 \pi^3 \rho_\text{tot}(t_\text{eva}))$, which is independent of $k$, the relevant integral is
\begin{widetext}
\begin{align}
\int_{\kappa_\text{eva}}^{\kappa_\text{i}} \text{d}\kappa \frac{T_\text{PBH}}{|\dot{\kappa}|} \frac{\kappa_\text{eva}\kappa^3}{\ee^\kappa -1} \simeq & \frac{T_\text{PBH,i}}{H_\text{eva}} \times \begin{cases}
\left( \frac{t_\text{eva}}{t_\text{eq,1}} \right)^{1/3} \kappa_\text{eva}^3 \left| \ln \kappa_\text{eq,1} \right|  & \propto k^3  \hspace{4.4mm}  \text{ for } k/a_\text{eq,1} \ll T_\text{PBH,i} \\
\frac{1}{2} \sqrt{\pi} \zeta (3/2) \kappa_\text{eva}^{5/2} & \propto k^{5/2} \hspace{2mm} \text{   for } k/a_\text{eva} \ll T_\text{PBH,i} \ll k/a_\text{eq,1} \\
48 \zeta (5) x_\text{eva}^{-1} & \propto k^{-1}  \hspace{2.6mm} \text{  for }  T_\text{PBH,i} \ll k/a_\text{eva}
\end{cases},
\end{align}
\end{widetext}
where $\kappa_\text{eva} \equiv k/(a_\text{eva} T_\text{PBH,i})$ and $\kappa_\text{eq,1} \equiv k/(a_\text{eq,1} T_\text{PBH,i})$, and $\zeta (\cdot )$ is the Riemann zeta function. 

These scaling properties have been confirmed by numerical integration (see the black dot-dashed line in Fig.~\ref{fig:gw_summary_appendix}).    Note that some parts of the scaling in the previous works are different from the above results, and we believe they are incorrect. 
Note also that, in Fig.~\ref{fig:gw_summary_appendix}, we take the high-frequency cutoff as the Planck scale at the evaporation time ($t_\text{eva}$) because there is no consensus on the Hawking radiation spectrum at the scales higher than the Planck one.  

The characteristic frequency is proportional to $T_\text{BH} \propto M_\text{PBH,i}^{-1}$, but the redshift factor depends on the reheating temperature, $a(T_\text{R}) \propto 1/T_\text{R} \propto M_\text{PBH,i}^{3/2}$.  Combining them, the scaling of the characteristic frequency is proportional to $M_\text{PBH,i}^{1/2}$. As long as the PBHs dominate the Universe, the height and the horizontal position of the $\Omega_\text{GW}$ curve hardly depend on $\beta$.

\subsection{GWs from mergers of binary PBHs}
This is separately studied in detail in the next appendix, but here we just summarize the spectral shape and the dependence of the peak wavenumber and peak height of $\Omega_\text{GW}$ on $M_\text{PBH,i}$ and $\beta$ for comparison.
Similarly to the other GWs contributions, the spectral index approaches $3$ on the largest scales.  The spectrum bends to the $k^{2/3}$ scaling~\cite{Ajith:2007kx, Ajith:2009bn}.  At which scale this happens is discussed in the next appendix.  The small-scale cutoff is given by the frequency of the GWs emitted just before the evaporation. 

The dependence of the characteristic scale is similar to the Hawking radiation case.
It is proportional to $M_\text{PBH,i}^{-1}$ at the source frame, but the redshift factor depends on $T_\text{R}^{-1} \propto M_\text{PBH,i}^{3/2}$.  Thus, the peak $k$ position scales as $M_\text{PBH,i}^{1/2}$.  The peak height of $\Omega_\text{GW}$ is proportional to $M_\text{PBH,i}^{6/37} \beta^{16/37}$.  This is valid only when the energy-density fraction of the non-relativistic matter in PBHs, $f_\text{PBH}$, is close to unity, which is always satisfied in the PBH-dominating scenario.


\section{\label{app:merger} Effects of mergers of PBHs}
In this appendix, we discuss the effects of formation of binary PBHs and their mergers.   If the merger event rate is substantial, it affects the mass distribution function of PBHs and may affect the dynamics of the reheating.  We will see that the energy density of the merged components can be (marginally) subdominant depending on  the initial PBH abundance, so it may or may not affect our estimation of $\Omega_\text{GW}$ induced just after the PBH evaporation. 

We also study the stochastic GW background from superposition of many PBH merger events. The spectrum is known to have a slope $\Omega_\text{GW}(k) \propto k^{2/3}$~\cite{Ajith:2007kx, Ajith:2009bn}.  Even though the typical frequency of GWs associated to the merger events is much higher than the typical frequency of the GWs induced by scalar modes just after the evaporation, the gentle slope of the spectrum, $k^{2/3}$, implies it might be observationally relevant.  In this context, we also have to take into account the validity of the extrapolation of this scaling.  We revisit the infrared (IR) part of the spectrum, which should reproduce the universal (white-noise) $k^3$ scaling~\cite{Cai:2019cdl}.   It turns out that such an IR modification is indeed relevant when we consider the GWs originating from mergers, and the GWs tend to be a subdominant component compared to the GWs induced just after evaporation. On the other hand, the merger-related GWs have a relatively large magnitude on smaller scales.  They are one of the dominating GWs on these scales as shown by the black dashed line in Fig.~\ref{fig:gw_summary_appendix}.

Effects of mergers of  PBHs in the early Universe were discussed in Appendix A of Ref.~\cite{Hooper:2019gtx}, based on the binary formation mechanism of Ref.~\cite{Bird:2016dcv},  and the authors concluded that they are ineffective unless the cosmic temperature is very high: $T \gtrsim 10^{11} \, \text{GeV} \times (10^4 \, \text{g} / M_\text{PBH})^{3/4}$. 
  More recently, the authors of Ref.~\cite{Zagorac:2019ekv} discussed PBH mergers in the PBH-dominating scenario based on another binary formation mechanism~\cite{Nakamura:1997sm}, which is more efficient than that of Ref.~\cite{Bird:2016dcv}.  
Here, we calculate the GW spectrum from superposition of GWs emitted by many merger events in the very early Universe before the PBH evaporation, improving the treatment on the GW energy spectrum from each merger event in Ref.~\cite{Zagorac:2019ekv}.  Instead, we neglect the time dependence of the PBH mass and in particular the effects of accretion of the surrounding radiation into PBHs\footnote{
Effects of accretion change the PBH mass at most only by $14 \%$~\cite{Zagorac:2019ekv}.  It is also discussed in Appendix B of Ref.~\cite{Hooper:2019gtx}, and they concluded that they are negligible unless the cosmic temperature is very high: $T \gtrsim 10^{14} \, \text{GeV} \times (10^4 \, \text{g}/M_\text{PBH})^{1/2}$.
}.  Our analyses are complementary to those in Ref.~\cite{Zagorac:2019ekv}.

In Refs.~\cite{ Sasaki:2016jop, Sasaki:2018dmp}, the authors investigated the PBH binary formation in the RD era.  The binary forms when the binary system is decoupled from the Hubble flow. The head-on collision is avoided because of the torque from the PBH closest to the binary. 
In the present context, the binaries are supposed to form during the eRD era, not the standard RD era (in which PBHs have already evaporated). Also, the merger events can occur only before the evaporation time.  These changes can be readily implemented by replacing their matter-radiation equality time $t_{\text{eq}}$ and the current cosmic time $t_{0}$ with $t_{\text{eq},1}$ and $t_{\text{eva}}$ respectively. We should also replace the PBH fraction of dark matter with unity, $f_\text{PBH} = 1$. 
Following Refs.~\cite{Wang:2016ana, Wang:2019kaf} in the context of LIGO/Virgo events, the $\Omega_\text{GW}$ just after the evaporation is 
\begin{align}
\Omega_\text{GW}(t_\text{eva}, f) = \frac{f}{a(t_\text{eva})}\frac{n_{\text{PBH}} (t_{\text{eva}})}{ \rho_\text{tot} (t_\text{eva})} \int_{t_\text{i}}^{t_\text{eva}} \dd t \frac{\dd P}{\dd t}  \frac{\text{d} E_\text{s}}{\text{d} f_\text{s}} (f_\text{s}),  \label{Omega_GW_merger}
\end{align}
where 
$n_{\text{PBH}} (t)$ is a physical PBH number density at $t$ and  $\dd P$ stands for a probability for a given PBH to merge from $t$ to $t + \dd t$ (see Eq.~\eqref{eq:prob}), and $\text{d}E_\text{s}/\text{d}f_\text{s} \, (f_\text{s})$ is the energy spectrum emitted by a single merger event as a function of the frequency at the source frame $f_\text{s} = f / a(t)$, and the subscript s means the source frame.  

\subsection{Merger rate and merged fraction}

Let us estimate the energy density fraction of PBHs already merged at the evaporation time.  If it is subdominant, the effect on the sudden evaporation dynamics is negligible.
Hence we treat $M_\text{PBH}$ as a constant for simplicity.
The generalization to the time-dependent mass case is desirable, but it is beyond the scope of the paper.
Also, we assume that the binary is not disrupted by encountering other PBHs once the binary is formed. 
Under these assumptions, the energy density fraction of merged PBHs at the evaporation time can be obtained from
\begin{align}
\Omega_\text{merged PBHs}(t_\text{eva}) = \frac{2 M_\text{PBH} n_{\text{PBH}} (t_{\text{eva}})}{ \rho_\text{tot} (t_\text{eva})} \int_{t_\text{i}}^{t_\text{eva}} \dd t \frac{\dd P}{\dd t}.
\label{eq:Omega_merged}
\end{align}
This probability $\text{d}P/\text{d}t$ is computed in Refs.~\cite{Sasaki:2016jop, Sasaki:2018dmp}. 
After appropriately replacing their matter-radiation equality time $t_{\text{eq}}$ and the current cosmic time $t_{0}$ with $t_{\text{eq},1}$ and $t_{\text{eva}}$ respectively, we find
\begin{align}
	\frac{\dd P}{\dd t} =  \frac{3}{58 t} \left( \frac{t}{T} \right)^{3/8} \left[  \frac{1}{ \left(1- \min \left[ \enn, \efinal \right]^2 \right)^{29/16}} -1 \right], 
\label{eq:prob}
\end{align}
where 
\begin{align}
	T \equiv \frac{3 \left( 8\pi M_\text{Pl}^{2} \right)^3 }{170 M_\text{PBH}^3} \left( \frac{4\pi f_\text{PBH}}{3} \right)^{-4} \left( \frac{4 \pi n_\text{PBH} (t_{\text{eq},1})  }{3} \right)^{-4/3},
\end{align}
is the time by which all the binaries merge. 
$f_\text{PBH}$ is the energy density fraction of the non-relativistic matter in PBHs, which is equal to unity in the PBH-dominating scenario but retained for generality.  The maximum eccentricities $\enn$ and $\efinal$ are given by
\begin{align}
1 - \enn^2 =& \left(  \frac{t}{T} \right)^{6/37}, \label{e_top} \\
1 - \efinal^2 
=&  \left( \frac{4\pi f_\text{PBH}}{3} \right)^{-32/21} \left(  \frac{t}{T} \right)^{2/7}. \label{e_side} 
\end{align}
These values $\enn$ and $\efinal$ are determined by the cutoff distance from the binary to the nearest neighbor PBH and by the largest possible distance between the PBHs in the binary, which is realized for the final binaries formed at the equality time, respectively. 
When $f_\text{PBH} =1$, the latter one is irrelevant since $\efinal > \enn$.

Note, however, that the applicability of the analytic formula of the merger probability to the $f_\text{PBH} \approx 1$ case is questioned in Ref.~\cite{Raidal:2018bbj} where discrepancies between the analytic estimate and $N$-body numerical simulations were found.  The simulations tell us that the binaries tend to be disrupted by an $N$-body cluster of surrounding PBHs (formed by Poisson fluctuations) and lose eccentricity.  This implies that the binary lifetime becomes longer and that the merger rate is suppressed.  It is not clear to us whether this also applies to the tiny PBH cases since the investigated mass ranges are different in 30 orders of magnitude.

By substituting Eq.~\eqref{eq:prob} into Eq.~\eqref{eq:Omega_merged}, we get the merged fraction just before evaporation~\cite{Zagorac:2019ekv}
\begin{align}
F_\text{merged} \equiv & \frac{\Omega_\text{merged PBHs}}{\Omega_\text{total PBHs}} \nonumber \\
\simeq &\begin{cases}
 \frac{1}{29} \left( 37 \left( \frac{t_\text{eva}}{T} \right)^{\frac{3}{37}} - 8 \left( \frac{t_\text{eva}}{T} \right)^{\frac{3}{8}} \right) & (t_\text{eva} < T) \\
 1 & (t_\text{eva} \geq T)
 \end{cases}.
\end{align}
The ratio $t_\text{eva} /T$ is evaluated as
\begin{align}
\frac{t_\text{eva}}{T} = 1 \times 10^{-17}  \left( \frac{f_\text{PBH}}{1} \right)^4   \left( \frac{M_\text{PBH,i}}{10^4 \, \text{g}} \right)^2  \left( \frac{\beta}{10^{-7}} \right)^{16/3}.
\end{align}
For the fiducial values $f_\text{PBH} = 1$ and $M_\text{PBH,i} = 10^4$~g, $F_\text{merged} = 5\%, 14\%$, and  $39\%$ if we take $\beta = 10^{-7}, 10^{-6}$, and  $10^{-5}$, respectively.  

The mass of once merged PBHs is approximately $2 M_\text{PBH,i}$.  The lifetime of these PBHs is $2^3 = 8$ times longer than the original PBHs.  By the time of their evaporation, the scale factor increases by a factor of $\sqrt{8}$.  If the merged fraction is less than $1/\sqrt{8}  \approx 35 \%$, the merged PBHs evaporate before they dominate the energy density and therefore they are harmless.  In this sense, $\beta \sim 10^{-5}$ is roughly the critical abundance for which the effect of the merger is significant.
We stress again that this estimation is based on the assumption that the binary is never disrupted after their formation until their merger, which is questioned in Ref.~\cite{Raidal:2018bbj} for $f_\text{PBH} =1$ and for $M_\text{PBH,i} = \mathcal{O}(10) M_\odot$.

\subsection{Revisiting GWs from merger events with ``IR cutoff''}

As mentioned at the beginning of this appendix and in Ref.~\cite{Cai:2019cdl}, we have to modify the spectrum of the merger-related GWs at the IR side.  
The energy spectrum from the inspiral-merger-ringdown process in the case of non-spinning BHs is obtained in Refs.~\cite{Ajith:2007kx, Ajith:2009bn}, and we augment it with an IR cutoff $f_\text{IR}$ as follows:
\begin{align}
\frac{\text{d}E_\text{s}}{\text{d}f_\text{s}}(f_\text{s}) = \frac{M_\text{c}^{5/3}}{12 M_{\text{Pl}}^{4/3} }  \times \begin{cases}
w_0 f_\text{s}^{2} & (f_\text{s} < f_\text{IR}) \\
f_\text{s}^{-1/3}  &   (f_\text{IR} \leq f_\text{s} < f_1 ) \\
w_1 f_\text{s}^{2/3} & (f_1 \leq f_\text{s} \leq f_2) \\
w_2 \frac{f_\text{s}^2}{\left( 1 + \frac{4 \left(f_\text{s}-f_2\right)^2}{\sigma^2} \right)^2} & (f_2 \leq f_\text{s} < f_3) \\
0 & (f_3 \leq f_\text{s}) 
\end{cases}, \label{inspiral-merger-ringdown}
\end{align}
where $w_0$, $w_1$ and $w_2$ are coefficients that make the spectrum continuous at $f_\text{s} = f_\text{IR}$, $f_1$ and $f_2$, respectively, and $M_\text{c}^{5/3} \equiv m_1 m_2 (m_1 + m_2)^{-1/3}$ is the chirp mass where we take $m_1 = m_2 = M_\text{PBH}$. Explicitly,  $w_0 = f_\text{IR}^{-7/3}$, $w_1 = f_1^{-1}$ and $w_2 = f_1^{-1} f_2^{-4/3}$.  These frequencies are given as $M_\text{PBH} f_1/(4 M_\text{Pl}^2) = 0.1125$,  $M_\text{PBH} f_2/(4 M_\text{Pl}^2) = 0.2565$, $M_\text{PBH} f_3/(4 M_\text{Pl}^2) = 0.3503$, and $M_\text{PBH} \sigma/(4 M_\text{Pl}^2) = 0.05952$.  See Ref.~\cite{Ajith:2009bn} for more general expressions in the case of non-equal masses and/or nonzero spins.

The IR cutoff of the frequency (in the source frame) should correspond to the largest possible orbit of the binary~\cite{Cai:2019cdl},
\begin{align}
f_\text{IR} =& \frac{1}{2\pi} \times 2 \sqrt{\frac{M_\text{PBH}}{8\pi M_{\text{Pl}}^{2}r^3}} , 
\end{align}
where $r$ is the radius of the orbit of the binary, which corresponds to $H^{-1}$ evaluated at the binary formation time.  
More precisely, this radius should be the length associated to the GW emission, and it should be shorter than $H^{-1}$ when the eccentricity $e$ is nonzero. We take it as the distance between a periastron and the closest focus (the center of gravity), $H^{-1} (1-e)$, which is (a half of) the shortest distance between the two PBHs in the binary trajectory.

Note that our estimation on the IR cutoff is conservative for the estimation on the magnitude of the GWs, which means that our $f_\text{IR}$ is probably larger than the true cutoff.  This is different from the estimation of the IR cutoff in Ref.~\cite{Cai:2019cdl}, where the authors presented a conservative estimate on the applicability of the universal IR cutoff scale, which means that their $f_\text{IR}$ is probably smaller than the true cutoff.

 The binary formation time (decoupling time) is estimated as $a_\text{dec}/a_\text{eq,1} = (\cmd/\cmd_\text{mean})^3$ where $\cmd$ is the comoving distance between the PBHs (we have used $\cmd_\text{max} = f_\text{PBH}^{1/3} \cmd_\text{mean} )$~\cite{Sasaki:2016jop, Sasaki:2018dmp}.  (Note the clash of notation: $\cmd$ here should not be confused with $x \equiv k \eta$ in the main text.) 
The cutoff can be rewritten as
\begin{align}
f_\text{IR} =& 
f_0 \left( \frac{\sma}{\sma_\text{max}} \right)^{-9/4} \left( 1 - e \right)^{-3/2}, \\
f_0 \equiv  & \frac{1}{\left(2\pi\right)^{3/2}}\sqrt{\frac{H_\text{eq,1}^3 M_\text{PBH}}{M_\text{Pl}^2}} ,
\end{align}
where $\sma = a_\text{dec} \cmd = a_\text{eq,1} \cmd (\cmd/\cmd_\text{max})^3$ 
with $\cmd_\text{max} = (f_\text{PBH}  / n_\text{PBH} (t) a^{3} (t))^{1/3}$ is the semi-major axis of the ellipse of the binary trajectory~\cite{Sasaki:2016jop, Sasaki:2018dmp} and $\sma_\text{max} = a_\text{eq,1} \cmd_\text{max}$.

The cutoff $f_\text{IR}$ as well as the energy spectrum depends on the parameters $\sma = \sma (t,e)$ and $e$ of the ellipse of the binary.  However, the probability in Eq.~\eqref{eq:prob} has been obtained by integrating the differential merger rate over $e$.  Therefore, we cannot use the equation directly, and we have to go back one step.
The differential probability for a given PBH binary to merge in the intervals of ($t$, $t + \dd t$) and ($e$, $e + \dd e$) is given by~\cite{Sasaki:2016jop, Sasaki:2018dmp}
\begin{align}
\frac{\dd^{2} P}{\dd e \dd t}= \frac{\pi^2 e f_\text{PBH}^{3/2} n_\text{PBH}^{1/2} (t_{\text{eq},1}) \sma^{3/2} }{3 \left(1 - e^2\right)^{3/2} t}  , \label{dP_formula}
\end{align}
where $\sma$ in terms of $t$ and $e$ is given by $\sma = (t/Q)^{1/4} / (1-e^2)^{7/8} $ with $Q = (3/170) (M_\text{PBH}/ 8 \pi M_{\text{Pl}}^{2})^{-3}$.
This relation $\sma = \sma (t,e)$ is obtained by inverting the relation between the lifetime of the binary $t$ in terms of the initial $\sma$ and $e$~\cite{Sasaki:2016jop, Sasaki:2018dmp}. Using Eq.~\eqref{dP_formula}, we can numerically compute the following integral to obtain the GW spectrum originating from the merger events:
\begin{align}
\Omega_\text{GW}(t_\text{eva}, f) =& \frac{f}{a(t_\text{eva})}\frac{n_{\text{PBH}}(t_{\text{eva}})}{ \rho_\text{tot} (t_\text{eva})} \int_{t_\text{i}}^{t_\text{eva}} \text{d} t \int_0^{e_\text{max}} \text{d} e \frac{\text{d}^2 P}{\text{d} t \text{d} e} \nonumber \\ 
&  \times   \frac{\text{d} E_\text{s}}{\text{d} f_\text{s}} (f_\text{s}) ,  \label{eq:Omega_GW_merger_integral-formula}
\end{align}
where $e_\text{max} \equiv \min \left[  \enn, \, \efinal  \right]$($= \enn$ in the PBH-dominating scenario).
The result is shown as the black dashed line in Fig.~\ref{fig:gw_summary_appendix}.
Note that the same method for taking into account the IR cutoff is applicable to the case of the binary non-evaporating PBH mergers in the present Universe by the replacement such as $t_\text{eva} \to t_0$ and $t_\text{eq,1} \to t_\text{eq}$.  

The dependence of the peak height on $M_\text{PBH,i}$ and $\beta$ comes from $\int \text{d} t  (\text{d}P/\text{d}t)$, which is approximated as
\begin{align}
\int^{t_\text{eva}}_{t_\text{i}} \frac{\text{d}t}{t} \left( \frac{t}{T} \right)^{3/37}  \sim \left( \frac{t_\text{eva}}{T} \right)^{3/37} \sim M_\text{PBH,i}^{6/37} \beta^{16/37} f_\text{PBH}^{12/37}.
\end{align}
Note that the IR cutoff does not affect the peak scale, so the integral over $e$ has been performed first.  There is no $\beta$ dependence from other parts in Eq.~\eqref{eq:Omega_GW_merger_integral-formula}, and the $M_\text{PBH,i}$ dependence from other parts cancels among them.  Note that the above expression is valid for the PBH-dominating scenario ($f_\text{PBH} = 1$), but when $f_\text{PBH} \ll 1$, $\efinal$ becomes relevant and the above expression is modified accordingly.

\section{Non-linear scale}
\label{app:nl_scale}
In this appendix, we explain how to derive the non-linear scale of the matter perturbations.
Since the analyses in this paper are based on the linear theory except for the interactions between the scalar and tensor perturbations, it is necessary to discuss the smallest scale that the analyses can be applied to.

\begin{figure}[ht] 
  \centering \includegraphics[width=1\columnwidth]{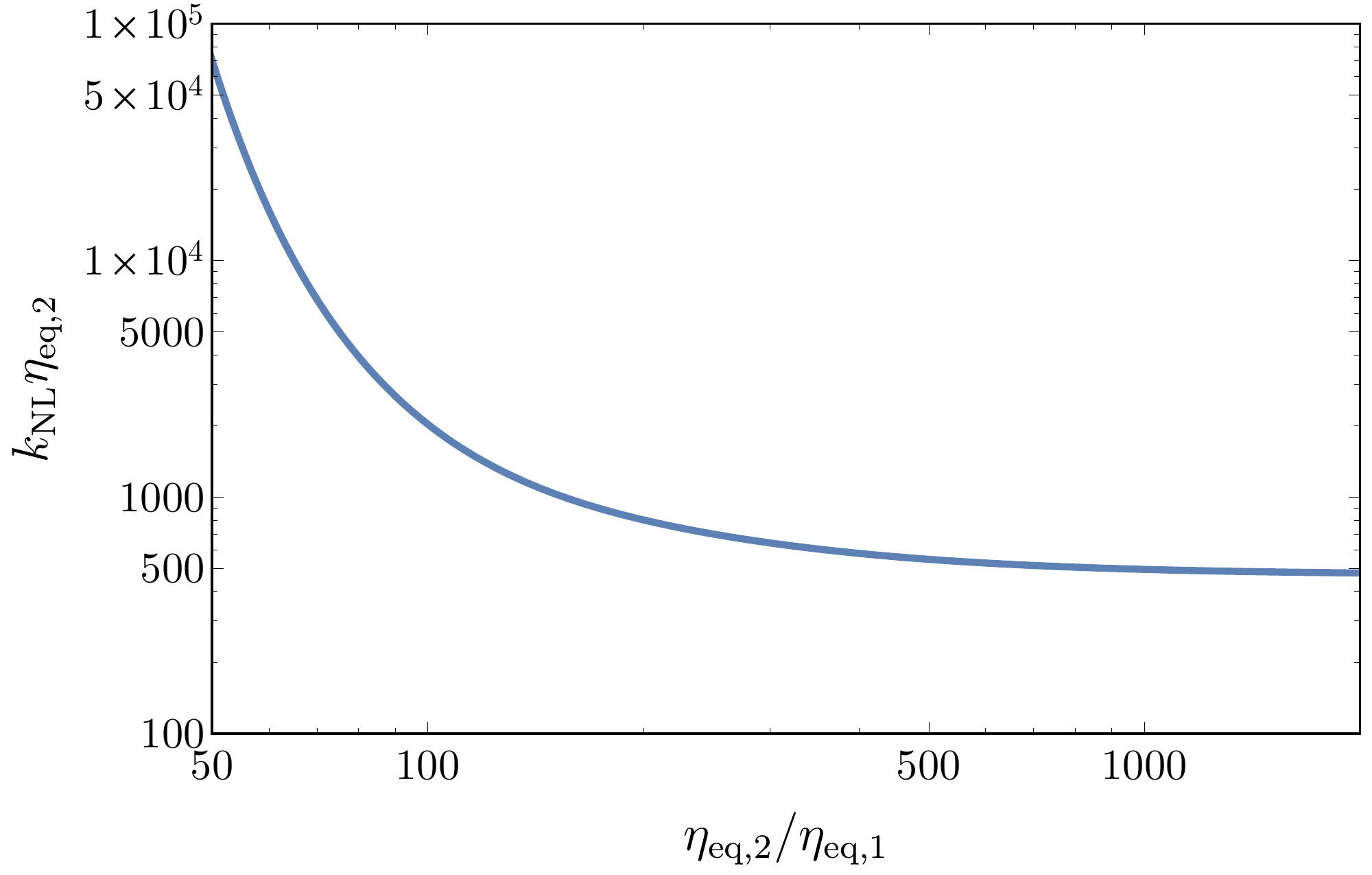}
        \caption{
        The relation between the wavenumber for the non-linear scale, defined in Eq.~(\ref{eq:k_nl_phi2}), and the length of the PBH-dominated era, characterized by $\eta_\eqs/\eta_\eqf$.
        }
        \label{fig:knl_ns1}
\end{figure}

During the PBH-dominated era, the gravitational potential is related to the matter perturbation on subhorizon scales as~\cite{Mukhanov:991646}
\begin{align}
  \frac{9}{10} k^2 \Phi_\text{plateau}(x_\eqf) \phi_k \simeq \frac{3}{2} \mathcal H^2 \delta_\text{m},
  \label{eq:phi_delta_m_rel}
\end{align}
where $\phi_k$ is the initial amplitude of the gravitational potential on superhorizon scales during the eRD era.\footnote{
  Since the $\Phi_\text{plateau}$ is normalized as $\Phi_\text{plateau}(x \rightarrow 0) \rightarrow 1$, the factor $9/10$ is multiplied in the left-hand side in Eq.~(\ref{eq:phi_delta_m_rel}).
   Note that $\Phi$ on superhorizon scales gets suppressed by the factor $9/10$ during the transition from the eRD era to the PBH-dominated era.
}
Using this equation, we estimate the non-linear scale $k_\text{NL}$, on which the perturbation becomes non-linear ($\delta_\text{m} = 1$) at the reheating.
Then, we can approximately derive the expression for $k_\text{NL}$ as 
\begin{align}
  \frac{3}{\sqrt{10}} k_\text{NL} \Phi_\text{plateau}^{1/2} (k_\text{NL} \eta_{\text{eq},1}) \phi_{k_\NL}^{1/2} \simeq \sqrt{\frac{3}{2}} \frac{2}{\eta_{\text{eq},2}}.
  \label{eq:k_nl_phi}
\end{align}
$\phi_k$ is related to the curvature perturbation as $|\phi_k| \simeq \frac{2}{3} |\zeta| \sim \frac{2}{3} \mathcal P_\zeta^{1/2}$.
Then, we can rewrite Eq.~(\ref{eq:k_nl_phi}) as
\begin{align}
  k_\text{NL} \Phi_\text{plateau}^{1/2} (k_\text{NL} \eta_{\text{eq},1}) \mathcal P_\zeta^{1/4}(k_\text{NL}) \simeq \sqrt{\frac{5}{2}} \frac{2}{\eta_{\text{eq},2}}.
  \label{eq:k_nl_phi2}
\end{align}
When we derive the cutoff scale $k_\text{NL}$, we substitute $2.1\times 10^{-9}$ into $\mathcal P_\zeta(k)$~\cite{Aghanim:2018eyx} throughout this paper for simplicity, which is a conservative assumption.\footnote{
  Strictly speaking, if we take into account the tilt of the power spectrum, such as $n_\text{s} = 0.96$, the $k_\text{NL}$ could become a little larger, which leads to the larger induced GWs.
}
Figure~\ref{fig:knl_ns1} shows the relation between $k_\NL \eta_\eqs$ and $\eta_\eqs/\eta_\eqf$.
Since the growth of matter perturbation during the eRD era is slower than that during the PBH-dominated era, the short duration of the PBH-dominated era leads to the large $k_\text{NL}$.\footnote{
In the limit of the large $\eta_\eqs/\eta_\eqf$, $k_\text{NL} \eta_\eqs$ asymptotes to a value about $470$ (see Fig.~\ref{fig:knl_ns1}). 
In this case, the relation between $k_\text{NL}$ and $M_\text{PBH,i}$ is given as $k_\text{NL} \simeq 1.1\times 10^{14}\, \text{Mpc}^{-1} (M_\text{PBH,i}/10^4\,\text{g})^{-3/2}$, where we have used Eqs.~(\ref{eq:tr_pbh_mass}) and (\ref{eq:kr_tr}).}

\section{Approximation formula for the induced GWs} 
\label{app:app_formula}

In this appendix, we explain the derivation of Eqs.~(\ref{eq:omega_gw_res_ana}) and (\ref{eq:omega_gw_res_w_sigma}).
In particular, we focus on the contribution from the resonance effect associated with the oscillations of the gravitational potential, which is present only in the case of GW production in the RD era~\cite{Ananda:2006af, Inomata:2019ivs}.
This contribution gives rise to a peak of the induced GW spectrum, which is relevant for constraints on the PBH abundance, and Eq.~(\ref{eq:omega_gw_res_ana}) represents such a contribution.
We emphasize again that our GW production mechanism occurs not during Hawking evaporation but just after the PBH evaporation so that the energy density of the Universe is dominated by radiation.

\subsection{Case of monochromatic mass function}

First, we explain the derivation of Eq.~(\ref{eq:omega_gw_res_ana}), which is the approximation formula in the case of the monochromatic mass function.
The resonance condition means the energy conservation ($|\vec{k}_1|/\sqrt{3} + |\vec{k}_2| /\sqrt{3} = |\vec{k}|$) under the condition of the momentum conservation ($\vec{k}_1 + \vec{k}_2 = \vec{k}$) where $\vec{k}_1$, $\vec{k}_2$, and $\vec{k}$ are the momenta of the two scalar source modes and the GW mode, respectively.  The factor $1/\sqrt{3}$ represents the sound speed in the thermal bath, while the speed of GW is $1$.
This is satisfied in a part of the integration region in Eq.~\eqref{eq:p_h_formula} where $u+v = \sqrt{3}$. 
For this resonant contribution, the oscillation average of the square of the function $I$ in Eq.~\eqref{eq:p_h_formula} in the case of a sudden reheating transition is given by Eq.~(B6) of Ref.~\cite{Inomata:2019ivs} multiplied with $(x - x_\text{eva}/2)^{-2}$.   As a counterpart of Eq.~(B7) in Ref.~\cite{Inomata:2019ivs}, we obtain
\begin{align}
\Omega_\text{GW}^{\text{(res)}} (\eta_\text{c} ,k) \simeq & \int_{-s_0(k/k_\text{NL})}^{s_0(k/k_\text{NL})} \text{d} s  \, \frac{3\sqrt{3} \, \mathcal{C} \left(1-s^2\right)^2}{40960000}  x_\text{eva}^7 \times  \nonumber \\ 
& \mathcal{P}_\zeta^\text{(eff)} \left( \frac{\left(\sqrt{3}+s\right)k}{2} \right) \mathcal{P}_\zeta^\text{(eff)} \left( \frac{\left(\sqrt{3}-s\right)k}{2} \right), \label{eq:omega_gw_res_definition}
\end{align}
where $\mathcal{C} \equiv 2 \int_{0}^{1} \text{d} y \text{Ci}(y)^2 \approx 2.3  
$ is a numerical constant, $\text{Ci}(y) \equiv - \int_y^\infty \text{d}y \cos (y) /y$ is the cosine integral function, and $\mathcal{P}_\zeta^\text{(eff)} (k) \equiv S^2 (k) \mathcal{P}_\zeta (k)$ is the effective power spectrum taking into account the suppression of the gravitational potential (see the discussion above Eq.~\eqref{eq:Phi_norm}). 
For a generic wavenumber $k$, the integration boundary $s_0(k/k_\text{NL})$ is $1$, but when it approaches the cutoff $k_\text{NL}$, it changes as Eq.~\eqref{eq:s0_def}.  We take into account this effect later, and tentatively set $s_0 = 1$.

Now, we introduce an approximation to Eq.~\eqref{eq:omega_gw_res_definition}.
The integration with respect to $s$ from $-1$ to $1$ corresponds to only an $\mathcal{O}(1)$ change of the argument of the power spectrum $\mathcal{P}_\zeta^{\text{(eff)}}$.  This implies that for any given value of $k$ and for a sufficiently smooth $\mathcal{P}_\zeta^{\text{(eff)}} (k)$, one can approximate the latter as a single power law spectrum with an effective tilt $n_\text{s,eff}(k)$.  The $k$-dependence of Eqs.~\eqref{eq:Phi_plateau} and \eqref{eq:s_low} tells us that $n_\text{s,eff} \simeq n_\text{s}$, $ n_\text{s}-2/3$, and $ n_\text{s} - 14/3$ up to a logarithmic correction, for $k \ll k_\text{eq,2}$, $k_\text{eq,2} \ll k \ll k_\text{eq,1}$, and $k \gg k_\text{eq,1}$, respectively.  Since the spectral shape of the resonance contribution extends to a large $k$ region, we take $n_\text{s,eff} = n_\text{s} - 14/3$ irrespective of the value of $k$, which is supported by the numerical calculations (see Fig.~\ref{fig:gw_spectrum_norm}).
This allows the analytic integration with respect to $s$,\footnote{
The naive interpretation of the $k^7$ dependence up to $\mathcal P(k)^2 S(k)^4$ in Eq.~(\ref{eq:omega_gw_res_appendix}) is as follows.
The factor $k^2$ arises in $f$ given in Eq.~(\ref{eq:f_def}) because the dominant term is proportional to $\Phi'\Phi'$, which gives the factor $k^4$ to $\mathcal P_h (\propto f^2)$.
Another factor arises from the integrals appearing in the expression of $\mathcal P_h$.
Eq.~(\ref{eq:p_h_formula}) includes the integrals over $v$ and $u$ and, before changing the variables, they are originally over the wavenumbers of two scalar modes, $|\vec k_1|$ and $|\vec k_2|$.
Since the resonance condition is given as $|\vec k_1|/\sqrt{3} + |\vec k_2|/\sqrt{3} = |\vec k|$, the two integrals gives only a single power of $k$.
From these observation, we can see that the power spectrum of the tensor perturbations are $\mathcal P_h \propto k^5$.
In addition, the factor $k^2$ comes from the fact that the energy density parameter of the tensor perturbations is proportional to the square of the time derivative of the tensor perturbations as $\rho_\text{GW} \propto \vev{h'h'} \propto k^2 \mathcal P_h$ (see Eq.~\eqref{eq:gw_formula}). 
Therefore, the energy density parameter has the $k^7$ dependence around the peak.
}
\begin{align}
\Omega_\text{GW}^{\text{(res)}} (\eta_\text{c} ,k ) \simeq & 
 2.9  \times 10^{-7} F(n_\text{s,eff}) E(k/k_\text{NL}) \nonumber \\
 & \times \left(k\eta_\text{eva}\right)^7  \mathcal{P}_\zeta^2 (k) S^4 (k) ,  \label{eq:omega_gw_res_appendix}
\end{align}
where $F(n_\text{s,eff})$ is defined with the hypergeometric function $_2 F_1$ as
\begin{align}
  F(n_{\text{s,eff}}) &\equiv \int^1_{-1}\dd s \left(1-s^2\right)^2 \left( \frac{3-s^2}{4} \right)^{n_\text{s,eff} -1} \nonumber \\
  & = \frac{3^{n_\text{s,eff} -1} 4^{2-n_\text{s,eff}}}{n_\text{s,eff} \left(3+2 n_\text{s,eff}\right)} \left( \left(\frac{2}{3} \right)^{n_\text{s,eff}} \left(n_\text{s,eff} -3\right) \right.\nonumber \\
  & \left. \quad + \left(n_\text{s,eff}^2 - n_\text{s,eff} +3\right) {}_2 F_1 \left( \frac{1}{2},-n_\text{s,eff};\frac{3}{2};\frac{1}{3} \right) \right), \label{eq:F(ns_eff)}
\end{align}
and $E(k/k_\text{NL})$ (satisfying $E = 1$ for $k \ll k_\text{NL}$) is a function introduced to take into account the reduced integral region $s_0 \neq 1$ around $k \simeq k_\text{NL}$, on which we discuss in the following.
 Since we already made approximations to obtain the above result, we do not aim to obtain the exact formula.  (It is anyway not meaningful to discuss the precise shape of the spectrum when $k$ becomes close to the non-linear scale.)  In the case of the sudden reheating transition with a power-law spectrum, the $s_0$-dependence has been calculated in Eq.~(B13) of Ref.~\cite{Inomata:2019ivs}.  Taking the ratio between the equation and Eq.~\eqref{eq:omega_gw_res_appendix}, we can extract the falling-off behavior, which is nothing but $E(k/k_\text{NL})$.   The expression becomes significantly simpler when we set $n_\text{s} =1$ for this ratio (this does not mean we set $n_\text{s,eff}=1$ or $n_\text{s}=1$ in Eq.~\eqref{eq:omega_gw_res_appendix}),
 \begin{align}
\label{eq:E_expression}  
 E(k/k_\text{NL}) = \frac{s_0(k/k_\text{NL})}{8} \left(15 - 10 s_0^2(k/k_\text{NL}) + 3 s_0^4(k/k_\text{NL})\right).
 \end{align}  
Since $s_0$ is defined as Eq.~(\ref{eq:s0_def}), it satisfies $E(k/k_\text{NL}) = 1$ for $k/k_\text{NL} \leq \frac{2}{1+\sqrt{3}}$ and $E(k/k_\text{NL}) = 0$ for $\frac{2}{\sqrt{3}} \leq k/k_\text{NL} $, and it is smoothly interpolated between the two regimes.

Substituting the numerical value of $F(0.96-14/3)\simeq 5.5$ and the expression of $E(k/k_\NL)$ given in Eq.~(\ref{eq:E_expression}) into Eq.~(\ref{eq:omega_gw_res_appendix}), we obtain Eq.~(\ref{eq:omega_gw_res_ana}).

\subsection{Effects of finite width}

Here, we generalize the approximation formula to the case of the mass function with finite width and derive Eq.~(\ref{eq:omega_gw_res_w_sigma}).
In the presence of finite width, the normalization factor, Eq.~\eqref{eq:sup_factor_with_sigma}, has an exponential dependence on $k$.  This does not satisfy the smoothness assumption that led to $F(n_\text{s,eff})$ in Eq.~\eqref{eq:F(ns_eff)}.  It should be modified as follows:
\begin{align}
F(n_\text{s,eff} , \sigma) \equiv & \int_{-1}^{1} \text{d}s \, \left(1-s^2\right)^2 \left( \frac{3-s^2}{4} \right)^{n_\text{s,eff} -1} \exp \left( - z \left(3+s^2 \right) \right),  
\label{eq:F(ns_eff)_sigma}
\end{align}
where $z \equiv  (c \sigma k \eta_\text{eq,2})^2 $ with $c$ introduced below Eq.~\eqref{eq:sup_factor_with_sigma}.
The exact analytic integration is not easy, so let us take the factorized form, $F(n_\text{s,eff} , \sigma) \approx F(n_\text{s,eff}) G(z)$, as an approximate ansatz. When $n_\text{s,eff} = 1$, the integral can be analytically done
\begin{align}
G(z) = \frac{15 \ee^{-4 z}}{64 z^{5/2}}\left( 2 \sqrt{z}\left(2 z -3\right) + \left(4z^2 - 4z +3\right) \ee^z \sqrt{\pi} \text{Erf} \left(\sqrt{z}\right) \right).
\end{align}
Then, we obtain Eq.~(\ref{eq:omega_gw_res_w_sigma}).
The error of the approximation is within a factor of 2 in the parameter region which we are interested in (see Fig.~\ref{fig:gw_profile_with_sigma}). 
Note that there is an additional $\ee^z$ factor in $G(z)$ compared to the naive expectation $G(z) \sim S(k,\sigma)^4 \sim \ee^{-4z}$.

\small
\bibliographystyle{apsrev4-1}
\bibliography{pbh_eva}

\end{document}